\documentclass[aps,prd,twocolumn,superscriptaddress,nofootinbib,showpacs]{revtex4-2}
\usepackage[pass,letterpaper]{geometry}
\pdfoutput=1 
\usepackage[]{graphicx}
\usepackage{color}
\usepackage{latexsym,amsmath,amsfonts,amssymb,bm}
\usepackage{cancel}
\usepackage{booktabs}
\usepackage{verbatim}
\usepackage{hyperref}
\usepackage{adjustbox}
\def\dd{\textrm{d}}

\def\newbeta{\hat{\beta}}

\begin{document}

\title{Hubble tension as a window on the gravitation of the dark matter sector: Exploration of a family of models}

\author{Jean-Philippe Uzan}
\email{uzan@iap.fr}
\affiliation{Institut d'Astrophysique de Paris, CNRS UMR 7095,
Sorbonne Universit\'e, 98 bis Bd Arago, 75014 Paris, France}

\author{Cyril Pitrou}
\email{pitrou@iap.fr}
\affiliation{Institut d'Astrophysique de Paris, CNRS UMR 7095,
Sorbonne Universit\'e, 98 bis Bd Arago, 75014 Paris, France}

\date{\today}  
\begin{abstract}
A family of simple and minimal extensions of the standard cosmological $\Lambda$CDM model in which dark matter experiences an additional long-range scalar interaction is demonstrated to alleviate the long lasting Hubble-tension while letting primordial nucleosynthesis predictions unaffected and passing by construction all current local tests of general relativity. This article describes their theoretical formulation and their implications for dark matter. Then, it investigates their cosmological signatures, both at the background and perturbation levels. A detailed comparison to astrophysical data is performed to discuss their ability to fit existing data. A thorough discussion of the complementarity of the low- and high-redshift data and on their constraining power highlights how these models improve the predictions of the  $\Lambda$CDM model whatever the combination of datasets used and why they can potentially resolve the Hubble tension. Being fully predictive in any environment, they pave the way to a better understanding of gravity in the dark matter sector.
\end{abstract}
\maketitle

\section{Introduction}

With the increase of the volume and quality of astrophysical data, the standard $\Lambda$CDM cosmological model is witnessing new tensions~\cite{Peebles:2022akh} questioning its ability to describe all data. Among them,  the Hubble constant  tension has become one of the main anomalies. It arises from the discrepancy between  the model-dependent determination of  $H_0$ from the {\em Planck} analysis of the cosmic microwave background (CMB)  combined with baryonic oscillations (BAO) data and the Hubble diagram interpretation, in particular from the SH0ES experiment~\cite{Riess:2011yx,Riess:2016jrr}, that is almost independent of physical assumptions. The former leads to the value  $H_0=(67.49 \pm0.53)$~km/s/Mpc~\cite{Planck:2018vyg,ACT:2020gnv,SPT-3G:2021wgf} while the latter, using Cepheid-calibrated supernovae concludes that $H_0 = (73.04 \pm1.04)$~km/s/Mpc~\cite{Riess:2021jrx}. This trend is confirmed by the lensing time delays of quasars from the H0LiCOW experiment~\cite{Wong:2019kwg} which determines $H_0 = 73.3^{+1.7}_{-1.8}$~km/s/Mpc. This results in a $\simeq 5\sigma$ tension on $H_0$. Indeed both methods assume a standard $\Lambda$CDM cosmological model in particular to infer distances and physical scales.

In a very crude way, this tension arises from the fact that the value of $H_0$ from CMB+BA0 differs from the one obtained at low-$z$ from the Hubble diagram. Hence, it is usually formulated~\cite{Schoneberg:2021qvd} as a low/high-redshift tension since in first approximation, the key physical parameters at the background level are the comoving sound horizon
\begin{equation}\label{e.rs}
r_s=\frac{1}{H_0}\int_{z_*}^\infty \frac{c_s \dd z}{H(z)/H_0} 
\end{equation}
with $z_*\sim1088$ the redshift at recombination, and the comoving angular diameter distance\footnote{See \S~\ref{subsec3A} for definitions.}
\begin{equation}\label{e.rs+1}
 R_{\rm ang}(z)=\frac{1}{H_0}f_K\left[\int_0^z \frac{\dd z}{H(z)/H_0} \right],
\end{equation}
related to the angular (or luminosity) distance, $D_A(z)=R_{\rm ang}(z)/(1+z)$ or $D_L=(1+z)R_{\rm ang}(z)$. Their ratio fixes the physical angular scales $\theta_*$ of the acoustic peaks. The precision of CMB observations sets a strong constraint on this quantity from which one can get the early time estimations of $H_0$ given the knowledge of the baryon and dark matter (DM) energy densities. It follows that most of the arguments on the $H_0$ tension circle around the sound horizon with two main categories of models~\cite{DiValentino:2021izs,Schoneberg:2021qvd,Abdalla:2022yfr}. ``Late time models" modify the expansion history after recombination, increasing $H_0$ but keeping $r_s$  unchanged. While lowering the Hubble tension, these models remain in disagreement with the BAO + Pantheon~\cite{Knox:2019rjx,Arendse:2019hev} data. ``Early time models" modify the expansion history before recombination, e.g. through energy injection around the recombination, changing both $H_0$ and $r_s$ so as to have a lower sound horizon to allow for a larger $H_0$. Generically such an early dark energy (DE) slows down the clustering of DM and the baryon-photon fluid, suppressing power on small scales~\cite{Caldwell:2003vp,Doran:2005sn}, and was shown to alleviate but not resolve the Hubble tension~\cite{Aylor:2018drw,Lin:2021sfs}. This led Ref.~\cite{Beenakker:2021vff} to single out 7 assumptions that need to be broken in order to resolve the Hubble tension. While the relative statistical merits of the proposed theories have been compared~\cite{DiValentino:2021izs,Schoneberg:2021qvd}, it was pointed out~\cite{Jedamzik:2020zmd}  that models reducing $r_s$ can never fully resolve the Hubble tension, if they are expected to be also  in agreement with other lower redshifts cosmological datasets. 

As always in cosmology, when facing a tension, one must first question the analysis of the data~\cite{Colgain:2022nlb,Murakami:2023xuy}, the consistency of different datasets and the accuracy of the theoretical tools to interpret them. Then, one shall question the validity of the cosmological hypothesis, i.e. a change in the cosmological solutions with the same physical theory before arguing for new physics~\cite{Uzan:2006mf,Uzan:2009mx,Uzan:2016wji}. This article will question the physics subtending the $\Lambda$CDM model, but in the less-controlled sector of the theory, i.e. the dark matter sector for which we have no direct experimental control. Nevertheless, and as we will demonstrate, even if such a minimal and fully allowed extension is possible and systematically lowers the Hubble tension compared to the $\Lambda$CDM, it will drive us to compare the high and low redshift data sets to understand the origin of the remaining tension, requiring a new look at some datasets. While not the main purpose of this work, it will call for further and broader dedicated analysis. Concerning the cosmological hypothesis, so far, some models arguing for a violation of spatial homogeneity~\cite{Kasai:2019yqn} or isotropy~\cite{Akarsu:2019pwn} have been explored and the validity of the Friedmann-Lema\^{\i}tre (FL) metric on all scales  to accurately interpret small angular scales of thin beam observations~\cite{Fleury:2013uqa,Fleury:2013sna,Clarkson:2011br} have been questioned. None of them have led to a viable solution that would avoid the introduction of new physics.

Indeed,  the number of models with new physics have grown extensively exploring many properties of the dark energy (DE) and DM sectors; see Refs.~\cite{DiValentino:2021izs,Schoneberg:2021qvd,Abdalla:2022yfr,Khalife:2023qbu} for reviews. Many of them assume a form of DE or an interaction of DM with a new matter component and in particular a scalar field. Among them, a large class assumes that DM interacts with DE~\cite{DiValentino:2019jae,Gomez-Valent:2023jgn}, most of which are purely phenomenological in the sense that an interaction term, e.g. $Q(H,a,\rho_{\rm DE},\rho_{\rm DM},\ldots)$, is assumed without being implemented in a complete field theory, in order to modify the cosmic expansion history during some era. This has been shown to generically not alleviate the Hubble tension.  Similarly, several models~\cite{Liu:2023kce,vandeBruck:2022xbk,Archidiacono:2022iuu,Bottaro:2023wkd} consider that DM can decay to DE or dark radiation. They generically suffer  from the fact that they imprint a suppression of the matter power spectrum~\cite{Schoneberg:2021qvd}. To finish, building on the fact that  the Einstein-Boltzmann equations used to predict CMB+BAO is invariant~\cite{Rich:2015jla,Ge:2022qws}  under the rescaling transformations: $H\rightarrow \lambda H$, $G\rho\rightarrow \lambda^2G\rho$,$\sigma_T n_e\rightarrow\lambda\sigma_T n_e$ and $A_s\rightarrow A_s\lambda^{1-n_s}$, the idea to consider a varying gravitational constant was proposed~\cite{Begue:2017lcw} and implemented in different ways~\cite{Knox:2019rjx,DiValentino:2021izs,Abdalla:2022yfr} as a generalisation of old extended quintessence models~\cite{Uzan:1999ch,Riazuelo:2001mg}. Most of them face difficulties with either primordial nucleosynthesis (BBN) and/or local constraints on deviations from general relativity (GR). In order for the sound horizon, and thus CMB predictions, to remain unchanged,  one needs to also modify the Thomson cross-section $\sigma_T$, i.e. the non-gravitational physics which is highly constrained.  The idea to ``compensate" the variation of the gravitational constant by a variation of the fine structure constant and/or electron mass have been considered~\cite{Hart:2021kad}, showing it can alleviate the Hubble tension despite the CMB~\cite{Planck:2014ylh} and other constraints on the variation of constants~\cite{Uzan:2002vq} that shall indeed be included together with the cosmological observations. We note that most of these studies postulate a phenomenological law of evolution of the chosen constant at the background level, e.g. $c(z)$. Indeed this means that one chooses a ``solution'' to an unknown theory so that perturbations cannot be treated consistently and the model is not predictive in any other physical situation. Whatever the constant to be considered varying, one must identify a dynamical field~\cite{Uzan:2002vq,Uzan:2010pm,Ellis:2003pw} and treat consistently all its effects, cosmological and non-cosmological. Any model that is either not mathematically complete nor fully predictive shall be assigned a lower credence in model comparison since it represents only a phenomenological description and not a theory. This is usually not included in model comparisons~\cite{DiValentino:2021izs,Schoneberg:2021qvd}, as well as  the non-cosmological constraints, which our model will avoid {\em by construction}.\\
 
These investigations indicate that the most promising models have to focus on the DM sector and avoid to modify the visible sector of the standard model (SM) fields. This article proposes a new road by focusing on the properties of gravitation in the DM  sector which is loosely constrained. More precisely, we assume the existence of a long-range scalar force acting on DM alone similar to a fifth force in the DM sector, which shall better be called {\em dark second force} since DM is supposed to have no charge under the three known non-gravitational interactions. We construct a theory which is a  simple, natural and minimal extension of GR with the introduction of a massless scalar field coupled to DM only so that all matter fields are unaffected. The cosmological models relying on this theory are minimal extensions of the standard $\Lambda$CDM. Generically, they can be seen as a particular class of bi-scalar tensor (ST) theories~\cite{Damour:1990tw,Coc:2008yu,Fuzfa:2007sv,Mohapi:2015gua}, i.e. ST theories with two conformal couplings, one in the visible sector and the second in the dark sector. Those have however been constrained from their effects on BBN and deviations from GR in the Solar system that arise from the coupling to SM matter. To avoid such constraints, we restrict to the  subclass in which visible matter remains transparent to the scalar interaction. A similar idea was considered in  Ref.~\cite{Thomas:2022ucg} but the scalar field being a quintessence field impacted the dynamics only at low redshift and was not used to address the Hubble tension. In the context  of the  swampland conjecture~\cite{Agrawal:2019dlm} it was however argued that such a coupling could alleviate the Hubble tension. Here, we further assume that $\varphi$ is massless so that it is not a dark energy model, and this will turn out to be an important feature of our construction. 

These hypotheses have a series of advantages, namely (1) we start from a consistent and well-defined field theory so that it is fully predictive in all physical situations, not only in cosmology. Besides it ensures the background and perturbation dynamics are treated consistently. (2) It ensures that all existing tests on the deviation from GR and of the weak equivalence principle (WEP) are safe. In particular the latest results MICROSCOPE~\cite{MICROSCOPE:2022doy} that set strong constraints on any light dilaton models~\cite{Berge:2017ovy} as well as all the constraints from the variation of fundamental constants~\cite{Uzan:2002vq,Uzan:2010pm} will be satisfied  since they are all performed, so far, only with the SM matter.  When it turns to cosmology, it has further generic features:  (1) the scalar field will always be subdominant in the matter budget so that it does not change the expansion history directly. As such {\em it is not a dark energy component}.  (2)  In particular it will have no direct effect on the dynamics of the universe at low redshift, (3) no signature on all astrophysical tests of GR and (4) no effect on BBN abundances predictions~\cite{Pitrou:2018cgg,Pitrou:2019nub}. (5) Since the new interaction is mediated by a massless scalar, the conservation equations imply that $\rho_{\rm DM}$ transfers to $\rho_\varphi$. As will be shown for the best-fit solutions, the scalar field will behave as a pressureless fluid around equivalence and recombination and then with a higher equation of state, hence redshifting faster than radiation at low redshift. This is a key difference with dark radiation that scales as the inverse fourth power, that will ensure that it circumvents the effects on the matter power spectrum. (6) Deep in the radiation era, the field will be frozen and its evolution being triggered by the DM to radiation energy density ratio, its dynamics will occur naturally around the last scattering surface. (7) In an effective way, the scalar interaction modifies the redshift evolution of the DM energy density; this can be interpreted equivalently either as a variation of an effective dark sector gravitational constant or by the fact that a fraction of DM disappears from the matter budget between equivalence and recombination. (8) To finish, as a consequence and as we shall demonstrate, generically these models have the same sound horizon as the $\Lambda$CDM but with a higher $H_0$ at the expense of a lower $\Omega_{\rm D0}$. This translates in a younger universe, hence fitting the model building constraints defined in Ref.~\cite{Bernal:2021yli}. It follows that while sharing some characteristics with previous models proposed in the literature, none of the former proposals encapsulate all these features.\\

The article is organized as follows. Section~\ref{sec1} defines a fully consistent  theory in which the SM fields are subjected to GR while DM is subjected to a massless ST theory, i.e. it gives its action, field equations and then discusses the properties of gravity in the DM sector before defining the cosmological  models that will serve our investigations. This extension is minimal: the theory introduces 1, or eventually 2, new free parameters, to describe the coupling of the scalar field to DM and the cosmological models have either 2 or 3 new free parameters since there is a freedom on the initial conditions of the scalar field. This allows us to compute unambiguously cosmological predictions both at the background and perturbation levels. Then, Section~\ref{sec2} details the background dynamics in order to discuss physically the main features of these models. In particular, it shows that all models share generic properties, in particular to trigger the transition between equality and last scattering surface (LSS) and to have a sound horizon similar to the $\Lambda$CDM but with a lower $H_0$. The perturbation theory and its numerical implementation are detailed in Section~\ref{sec3} which then allows us to perform a full MCMC comparison to astrophysical data in Section~\ref{sec4} in order to discuss the power of this class of models to alleviate or eventually solve the Hubble tension. This leads us to conclude that this class of models do alleviate the Hubble tension. The properties of the best fits are discussed, in particular their generic predictions of a younger age for the universe. We summarize our findings in Section~\ref{secconlc} and discuss the limits and the potential systems to test the existence of a long range scalar interaction in the DM sector.

\section{Definition of the theory}\label{sec1}

This section defines the action of the theory and derives its equations of motion (\S~\ref{secII1}) before comparing the properties of gravity in the visible and dark sectors (\S~\ref{secII2}) and then defining the free functions that are used in our cosmological investigation (\S~\ref{secII3}).

\subsection{Action and field equations}\label{secII1}

\begin{figure}[tb]
 	\centering
 	\includegraphics[width=0.45\textwidth]{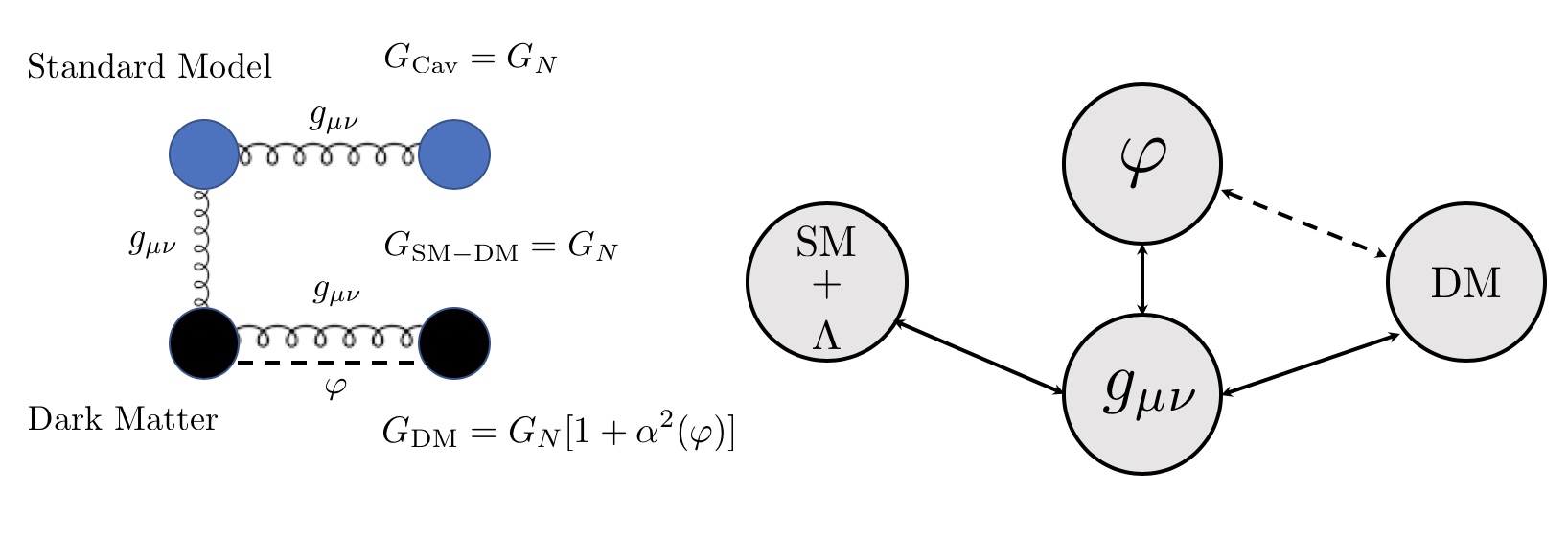}
 	\caption{Structure of the model. [{\it Left}] While the standard matter sector is governed by general relativity, the DM sector experiences an extra long range scalar interaction. As a consequence the strength of the gravitational interaction is different in the SM and DM sectors. Since SM fields are transparent to the scalar force, the gravitation between DM and SM fields is described by GR. The gravitational constant between to DM particles depends on the coupling $\alpha(\varphi)$ and is thus varying in space and time. [{\it Right}] The SM sector feels $\varphi$ and DM only through the effect of their stress-energy on the gravitational field $g_{\mu\nu}$ to which it is minimally coupled while DM particles experience an extra-scalar force with coupling $\alpha(\varphi)$.}
 	\label{fig:1}
 	\vspace{-0.25cm}
 \end{figure}

The action is decomposed as
\begin{equation} \label{e.TH1}
 S = S_{\rm GR} + S_{\rm SM} + S_{\varphi} + S_{\rm D}
\end{equation}
where we have introduced a new degree of freedom in the form of the scalar field\footnote{We use the standard normalisation used in scalar-tensor theories, see e.g. Ref.~\cite{Damour:1992we}.} $\varphi$; see Fig.~\ref{fig:1}. The action for the visible sector are
\begin{eqnarray}
 S_{\rm RG} &=&  \int \frac{\dd^4x}{16\pi G}\sqrt{-g}\left( R - 2\Lambda_0 \right)\label{e.LRG}\\ 
 S_{\rm SM}  &=&  \int \dd^4x \sqrt{-g} {\cal L}_{\rm SM}[\psi; g_{\mu\nu}] \label{e.LSM}
\end{eqnarray}
while the DM sector is modeled by
 \begin{eqnarray}
 S_{\varphi}  &=&- \int \frac{\dd^4x}{16\pi G}\sqrt{-g}\left[  2g^{\mu\nu} \partial_\mu\varphi\partial_\nu\varphi +  4V(\varphi)\right] \\
 S_{\rm D} &=& \int \dd^4x \sqrt{-\tilde g} {\cal L}_{\rm D}[\psi; \tilde g_{\mu\nu}]  \label{e.TH5}
\end{eqnarray}
with the DM-metric
\begin{equation}\label{e.tildeg}
\tilde g_{\mu\nu}=A^2(\varphi)g_{\mu\nu}.
\end{equation}

${\cal L}_{\rm SM}$ and ${\cal L}_{\rm D}$ are the Lagrangians of the standard model and dark matter sectors. The two free functions $V$ and $A$ describe the field potential and coupling to DM respectively. This theory is defined in the {\it SM-frame} which for DM corresponds to an Einstein frame. Since all observables concern SM fields we work in this frame in which nothing departs from GR for the SM fields with which all tests of GR have been performed so far. Indeed, the DM sector is expressed in its Einstein frame but all the usual notions of mass and energy, distance and time, etc. are defined in the SM-frame since all measurements are performed with SM fields, even for the DM masses and energy density since they act on SM fields only through $g_{\mu\nu}$ only. This is a difference with standard ST theories~\cite{Damour:1992we} (see appendix~\ref{appB} for a discussion on the frames).

The equations of motion deduced from the action are
\begin{eqnarray}
&& G_{\mu\nu}+\Lambda_0 g_{\mu\nu} =\,  \kappa T_{\mu\nu} + \kappa T^{(\varphi)}_{\mu\nu} + \kappa T^{\rm(DM)}_{\mu\nu} \label{FEq_1}\\
&& \nabla_\mu T^{\mu\nu} =0 \label{FEq_2}\\
&& \nabla_\mu T_{\rm(DM)}^{\mu\nu} =  \alpha(\varphi) T^{\rm(DM)}_{\sigma\rho} g^{\sigma\rho} \partial^\nu\varphi\label{FEq_3}  \\
&& \Box \varphi = \frac{\dd V}{\dd\varphi} - \frac{\kappa}{2} \alpha(\varphi) T^{\rm(DM)}_{\mu\nu} g^{\mu\nu}\label{FEq_4}
\end{eqnarray}
where we have defined $\kappa\equiv8\pi G$ and
\begin{equation}\label{e.defalphabeta}
\alpha(\varphi) = \frac{\dd \ln A}{\dd\varphi}, \qquad
 \newbeta(\varphi)\equiv \frac{\dd\alpha}{\dd\varphi}\,.
\end{equation}
The stress-energy tensors of visible and dark matters are
\begin{equation}\label{e.defT}
T^{\mu\nu} =\frac{2}{\sqrt{-g}}\frac{\delta {\cal L} \sqrt{-g}}{\delta g_{\mu\nu}}
\end{equation}
and the stress-energy tensor of the scalar field is 
\begin{eqnarray}
 \kappa T^{(\varphi)}_{\mu\nu}&=&2\partial_\mu\varphi\partial_\nu\varphi-\left[(\partial_\alpha\varphi)^2 + 2 V \right] g_{\mu\nu}\,.
\end{eqnarray}
It is easily checked that
\begin{equation}\label{e.DTDM}
\nabla_\mu\left[ T_{\rm(DM)}^{\mu\nu} + T_{(\varphi)}^{\mu\nu}\right]=0
\end{equation}
showing that the energy transfer occurs only, as designed, between $\varphi$ and DM, which is indeed nothing else but the action-reaction law. Note that the DM energy density measured by an observer comoving with $u^\mu$ [$g_{\mu\nu} u^\mu u^\nu=-1$] is $\rho_{\rm D}= T_{\mu\nu} u^\mu u^\nu$. Indeed, one could have considered the Einstein frame stress-energy tensor $\tilde T_{\mu\nu}$ defined as in Eq.~(\ref{e.defT}) but using the metric $\tilde g_{\mu\nu}$, so that $\tilde T_{\mu\nu}=A^2 T_{\mu\nu}$ and the associated energy density measured by an observer with 4-velocity $\tilde u^\mu$ satisfying $\tilde g_{\mu\nu} \tilde u^\mu \tilde u^\nu=-1$ would then be $\tilde\rho_{\rm D} =A^4(\varphi)\rho_{\rm D}$. This energy is not measurable. Since all observations and experiments are made with standard model fields, there is no possible ambiguity and the energy density inferred from them is $\rho_{\rm D}$ defined from $T_{\rm(DM)}^{\mu\nu}$. The metric $\tilde g$ is simply a convenient intermediary for implementing the scalar force but all physical interpretations are drawn from $g$.

Note that a series of works~\cite{Damour:1990tw,Coc:2008yu,Fuzfa:2007sv,Mohapi:2015gua}  have assumed the existence of a scalar interaction with different couplings between the visible and the dark sector in order to investigate the dark-visible equivalence principle. The coupling to SM fields induces, as already mentioned, a time variation of the Newton constant, modifications of the BBN abundances and deviations from GR in the Solar system so that they are strongly constrained. The present model evades these constraints by construction.

\subsection{Gravitation of the DM sector}\label{secII2}

Before studying the cosmology, let us consider the effect of the scalar-tensor theory on the motion of particles. In the SM sector, the matter fields couple to the metric $g_{\mu\nu}$ so that the equation of a point particle of mass $m$ and charge $q$ derives from the action
\begin{equation}\label{e.action}
S_{\rm pp}=-mc^2\int \sqrt{-g_{\mu\nu}u^\mu u^\nu}\dd\tau+ q \int {\cal A}_\mu u^\mu \dd\tau
\end{equation}
where $\tau$ is the proper time and $u^\mu$ the tangent vector to  the worldline, i.e.  $u^\mu=\dd X^\mu/\dd\tau$. It satisfies $u_\mu u^\mu =-c^2$ and ${\cal A}_\mu$ the potential vector. The variation of this action gives the usual equation of motion
\begin{eqnarray}\label{e.motionSM}
mc^2 \gamma^\mu = q {F^\mu}_\nu u^\nu
\end{eqnarray}
with the 4-acceleration $\gamma^\mu\equiv u^\nu\nabla_\nu u^\mu=\dd u^\mu/\dd\tau$ which satisfies $\gamma^\mu u_\mu=0$,  $F_{\mu\nu} = \partial_\mu  {\cal A}_\nu -\partial_\nu {\cal A}_\mu$ the Faraday tensor. For a neutral particle, this is indeed the equation for the geodesics of the metric $g$.

Now, in the DM sector, the matter fields couple to the metric $A^2(\varphi)g_{\mu\nu}$. The equation of a DM point particle of mass $m$ is
\begin{equation}\label{e.action2}
S_{\rm pp}=-c^2\int m A(\varphi) \sqrt{-g_{\mu\nu}u^\mu u^\nu}\dd\tau+ q \int {\cal A}_\mu u^\mu \dd\tau.
\end{equation}
Particles with $q=0$ follow geodesics of the metric $\tilde g_{\mu\nu}$, since the theory satisfies the weak equivalence principle within the DM sector. The equation of motion follows as
\begin{eqnarray}\label{e.motion}
mc^2 \gamma^\mu = \frac{q}{A(\phi)}{F^\mu}_\nu u^\nu - mc^2\frac{\partial \ln A}{\partial\phi}\perp^{\mu\nu}\nabla_\nu\varphi
\end{eqnarray}
with  $\perp_{\mu\nu}\equiv g_{\mu\nu}+u_\mu u_\nu/c^2$ the projector on the 3-space normal to $u^\mu$, which indeed ensures that $u^\mu u_\mu=-c^2$; see e.g. Ref.~\cite{Uzan:2010pm}. It follows that the fifth force,
\begin{equation}\label{e.5th}
F^\mu =  - mc^2\alpha(\varphi)\perp^{\mu\nu}\nabla_\nu\varphi,
\end{equation}
remains perpendicular to the 4-velocity, $u_\mu F^\mu=0$. \\

In the weak field, or Newtonian, limit $g_{00}=-(1+2\Phi_{\rm N}/c^2)$. Let us restrict to an uncharged particle, as expected for DM (the general case is described in Ref.~\cite{Uzan:2020aig} in case the DM enjoys a charge in the dark sector). First, clearly, in the Galilean limit the projector plays no role.  In 3-dimensional notations $X^\mu=(T,{\bm X})$, the 3-velocity and 3-acceleration are defines as
\begin{equation}
{\bm V}=\frac{\dd{\bm X}}{\dd T}, \quad
{\bm a}=\frac{\dd{\bm V}}{\dd T},
\end{equation}
with the convention ``${\bm V}=V^i$" with $i=1\ldots3$ (see Ref.~\cite{DUbook} for details).  For a non-relativistic particle, at first order in $v/c$, the equations of motion~(\ref{e.motion}-\ref{e.5th}) of a test particle, respectively of the SM and DM sectors, reduce to 
\begin{eqnarray}
{\bm a}&=& -\nabla \Phi_{\rm N},  \label{e.adyn} \\
{\bm a}_{\rm D}&=& -\nabla \Phi_{\rm N}  - \alpha \nabla \varphi \nonumber\\
  &\equiv& -\nabla [ \Phi_{\rm N} + \ln A(\varphi)]\equiv -\nabla\Phi_{\rm D}\label{e.adyn2}
\end{eqnarray}
so that the general expression of the effective DM gravitational potential is
\begin{equation}
\Phi_{\rm D} = \Phi_{\rm N}+ \ln A(\varphi).
\end{equation} 
Now, at lowest order in perturbations, $R_{00}=\Delta\Phi_N/c^2$. Rewriting Eq.~(\ref{FEq_1}) as $R_{\mu\nu}=\kappa (T_{\mu\nu}-Tg_{\mu\nu}/2)$, the Einstein equation for a dust fluid (or point particle) with stress-energy tensor $T_{\mu\nu}=\rho c^2 u_\mu u_\nu$ (so that $T=-\rho c^2$) gives $\Delta\Phi_N/c^2 = (8\pi G/c^4) \rho c^2/2$ that is the standard Poisson equation
\begin{equation}\label{e.poisson}
\Delta\Phi_{\rm N} = 4\pi G \rho.
\end{equation}
Now, assuming that $\varphi=\varphi_0+\delta\varphi$, the Klein-Gordon equation~(\ref{FEq_4}) gives
\begin{equation}\label{e.yukawa}
\left(\Delta -  m^2_\varphi\right)\delta\varphi= 4\pi G \alpha_0 \rho_{\rm D}.
\end{equation}
with $m^2_\varphi=V''(\varphi_0)$ and $\alpha_0=\alpha(\varphi_0)$. Working in units in which $c=1$, these two equations are solved in terms of the Green functions, i.e. with the source $\delta^{(3)}({\bm x}-{\bm x}')$,
\begin{equation}
{\cal G}_{m_\varphi}({\bm x}-{\bm x}') = -\frac{1}{4\pi}  \frac{\hbox{e}^{-m_\varphi |{\bm x}-{\bm x}'|}}{|{\bm x}-{\bm x}'|}
\end{equation}
as
\begin{eqnarray}
\Phi_{\rm N} &=& 4\pi G\int\dd^3{\bm x}' {\cal G}_{0}({\bm x}-{\bm x}') \rho({\bm x}'), \\
\delta\varphi  &=& 4\pi G\alpha_0 \int\dd^3{\bm x}' {\cal G}_{m_\varphi}({\bm x}-{\bm x}') \rho_{\rm D}({\bm x}'),
\end{eqnarray}
which for a point particle of mass $M$ gives
\begin{eqnarray}
\Phi_{\rm N} = \frac{GM}{r}, \quad
\delta\varphi  = \frac{GM_{\rm D}\alpha_0 \hbox{e}^{-m_\varphi r}}{r}.
\end{eqnarray}
Given that $A=A_0(1+\alpha_0\delta\varphi)$, Eq.~(\ref{e.adyn2}) simplifies to
\begin{eqnarray}
&&{\bm a}_{\rm D} = -\nabla \Phi_{\rm N}  - \alpha_0 \nabla \delta\varphi,
\end{eqnarray}
with
\begin{equation}\label{e.potD}
\Phi_{\rm D} = \frac{GM_{\rm D}(1+\alpha^2_0 \hbox{e}^{-m_\varphi r})}{r},
\end{equation}
with the Compton wavelength of the field $\lambda_\varphi=\hbar/m_\varphi c$. For a massless dilaton, a DM particle will experience a Newton-like force but with a modified gravitational constant,
 \begin{equation}\label{e.Gcav}
 G_{\rm DM}= G(1+\alpha_0^2),
 \end{equation}
which will be measured e.g. in a Cavendish experiment~\cite{Damour:1992we} between 2 DM massive spheres. The first term is the usual gravitational force, i.e. due to the exchange of a graviton while the second arises from the scalar fifth force, i.e. due to the exchange of a scalar. 

The theory satisfies the weak and strong equivalence principle in the visible sector and violates the strong equivalence principle in the DM sector. Indeed $G_{\rm DM}$ is dynamical and vary on cosmological scales. Obviously, it remains so far impossible to measure its value directly in the laboratory. In our local environment, the DM density is estimated~\cite{deSalas:2020hbh} to lie within $0.4$ and $0.6$~GeV/cm$^3$, too small a value to have observable dynamical effects; see however Ref.~\cite{Berge:2019zjj} for some proposed experiments. But this DM gravity in the DM sector may have some consequences on astrophysical scales, in particular for the structure of dark matter haloes and the relation between the standard and dark matters in galaxies.

\subsection{Definitions of the models}\label{secII3}

The previous theory depends on 2 arbitrary functions $V$ and $A$. In this article, we shall consider the following choices:
\begin{enumerate}
\item The family of minimal models introduced in our companion letter~\cite{PU_0} for a massless dilaton and quadratic interaction
\begin{equation}\label{e.defmodel1}
V(\varphi)=0,\qquad A(\varphi) =1+ \frac{1}{2}\beta\varphi^2,
\end{equation}
so that Eq.~(\ref{e.defalphabeta}) implies
\begin{equation}
\alpha(\varphi)= \frac{\beta \varphi}{1+ \frac{1}{2}\beta\varphi^2}, \quad
\newbeta(\varphi) = \beta  \frac{\left(1-\frac{1}{2}\beta \varphi^2\right)}{\left(1+ \frac{1}{2}\beta\varphi^2\right)^2}.
\end{equation}
The theory introduces a single new constant parameter, $\beta$.
\item The family of extended minimal models introduced for a massless dilaton introducing a quartic coupling
\begin{equation}\label{e.defmodel2}
V(\varphi)=0,\qquad A(\varphi) = 1+ \frac{\lambda}{4}\varphi^4.
\end{equation}
It introduces a single new constant parameters, $\lambda$.
\item $(\beta,\lambda)$ models as a combination of the two previous ones
\begin{equation}\label{e.defmodelE}
V(\varphi)=0,\qquad A(\varphi) = 1+ \frac{1}{2}\beta\varphi^2+\frac{\lambda}{4}\varphi^4.
\end{equation}
\item The family of quadratic coupling, as chosen in light-dilaton models~\cite{Damour:1990tw,Damour:1992we},
\begin{equation}
\alpha(\varphi)=\beta\varphi, \quad
\newbeta(\varphi) = \beta
\end{equation}
that derives from
\begin{equation}\label{e.defmodel4}
V(\varphi)=0,\qquad A(\varphi) = \hbox{e}^{\frac{1}{2}\beta\varphi^2},
\end{equation}
with one single new constant parameter $\beta$. Indeed it corresponds approximately to  a model~(\ref{e.defmodelE}) with $\lambda=\beta^2/2$.
\item The family of couplings inspired by axion physics~\cite{Sikivie:2006ni}
\begin{equation}\label{e.defmodel5}
V(\varphi)=0,\qquad
A(\varphi) = 1+\beta(1-\cos k\varphi)^n.
\end{equation}
At small $\varphi$, when $n=1$  it corresponds approximately to  a model~(\ref{e.defmodelE}) with $\beta=\beta k^2$ and $\lambda=-\beta k^4/6$.
\end{enumerate}
We shall refer to the first three models as $\Lambda\beta$CDM, $\Lambda\lambda$CDM and $\Lambda(\beta,\lambda)$CDM. The last two fall in the family $\Lambda(\beta,\lambda)$CDM with $\lambda$ of different signs and are better motivated physically. 

\subsection{Discussion}

The equations of \S~\ref{secII1} allow the investigation of any physical system, while the analysis in the Newtonian limit of \S~\ref{secII2} can allow for instance to implement the new interaction in Vlasov-Poisson equations or to describe DM haloes in the non-linear regime. Indeed DM and baryon will not fall in the same way which can bias the relation between their distributions. The subclass of models defined in \S~\ref{secII3} have one or two extra-parameters and all assume a vanishing potential. 

While all the models we consider assume $V=0$, we provide all the equations with a non-vanishing potential to allow for an easy generalisation such as non-minimal dark energy models. Note also that the distinction between matter and interaction is not obvious since all fields have an energy density that enters the Einstein equation. When this field is coupled to some matter fields, it is responsible for an interaction. In the case at hand, $\varphi$ is always negligible in the matter budget and its density is driven by $\rho_{\rm D}$, see Fig.~\ref{fig:2} below.

\section{Background cosmology}\label{sec2}

\subsection{Background equations}\label{subsec3A}

To discuss the cosmological implications of these models, let us consider a Friedmann-Lema\^{\i}tre spacetime with its usual metric
\begin{equation}
 \dd s^2 = -\dd t^2 + a^2(t)\gamma_{ij}\dd x^i \dd x^j
\end{equation}
where 
$$
\gamma_{ij}\dd x^i \dd x^j=\dd\chi^2+f_K^2(\chi)\dd\Omega^2
$$ 
is the spatial metric and $a$ the scale factor. $f_K(\chi)=\lbrace \sinh\chi,\chi,\sin\chi\rbrace$ respectively for $K=-1,0,+1$. The redshift of comoving observers and Hubble function are defined, assuming the convention $a_0=1$, as $1+z=1/a$ and $H=\dot\ln a$ with a dot referring to a derivative with respect to the cosmic time. 

At the background level, the Einstein equations give the Friedmann equations
\begin{align}
 & 3\left(H^2 + \frac{K}{a^2}\right) =    \kappa( \rho +   \rho_{\rm D}+\rho_\varphi)+ \Lambda \label{einsteinEF1}\\
  & -3 \frac{\ddot a}{a^2} = \frac{\kappa}{2} (\rho + 3P +  \rho_{\rm D}+3P_{\rm D}) -\Lambda
   + 2 \dot\varphi^2 - 2V  \label{einsteinEF2}
\end{align}
with the scalar field energy density and pressure,
\begin{equation}\label{def.rhophi}
 \kappa \rho_{\varphi} = \dot\varphi^2+2V,\quad
  \kappa P_{\varphi} = \dot\varphi^2-2V.
\end{equation}
While the conservation equations in the SM sector remain unchanged,
\begin{eqnarray}
&& \dot\rho_i + 3H (\rho_i +P_i)= 0, \label{e.c0}
\end{eqnarray}
for DM they become
\begin{eqnarray}
&& \dot\rho_{\rm D} + 3H (\rho_{\rm D} +P_{\rm D}) = \alpha (\varphi)(\rho_{\rm D} -3P_{\rm D})\dot \varphi\label{e.c2}
\end{eqnarray}
and the Klein-Gordon equation is
\begin{eqnarray}\label{kgEF}
 \ddot\varphi + 3H\dot\varphi &=& -\frac{\dd V}{\dd\varphi} -\frac{\kappa}{2} \alpha  (\rho_{\rm D}-3P_{\rm D}).
 \end{eqnarray}

As usual, we define the cosmological parameters
\begin{equation}
\Omega_{i 0}=\frac{8\pi G \rho_{i 0}}{3H_0^2},\quad
\Omega_{\Lambda 0}=\frac{\Lambda}{3H_0^2}, \quad
\Omega_{K 0}=-\frac{K}{3H_0^2}
\end{equation}
for baryons ($i=$~b), radiation ($i=$~r) and dark matter ($i=$~DM), the cosmological constant and the spatial curvature and we define the matter density parameter as
\begin{equation}
\Omega_{\rm m} \equiv \Omega_{\rm b0}+\Omega_{\rm D0}.
\end{equation}
Concerning the scalar field, we define
\begin{equation}
 \Omega_{\dot\varphi} =\frac{\dot\varphi^2}{3H_0^2}\,,
 \quad
 \Omega_V = \frac{2V}{3H_0^2}.
\end{equation}

\begin{figure}[htb]
 	\centering
 	\includegraphics[width=\columnwidth]{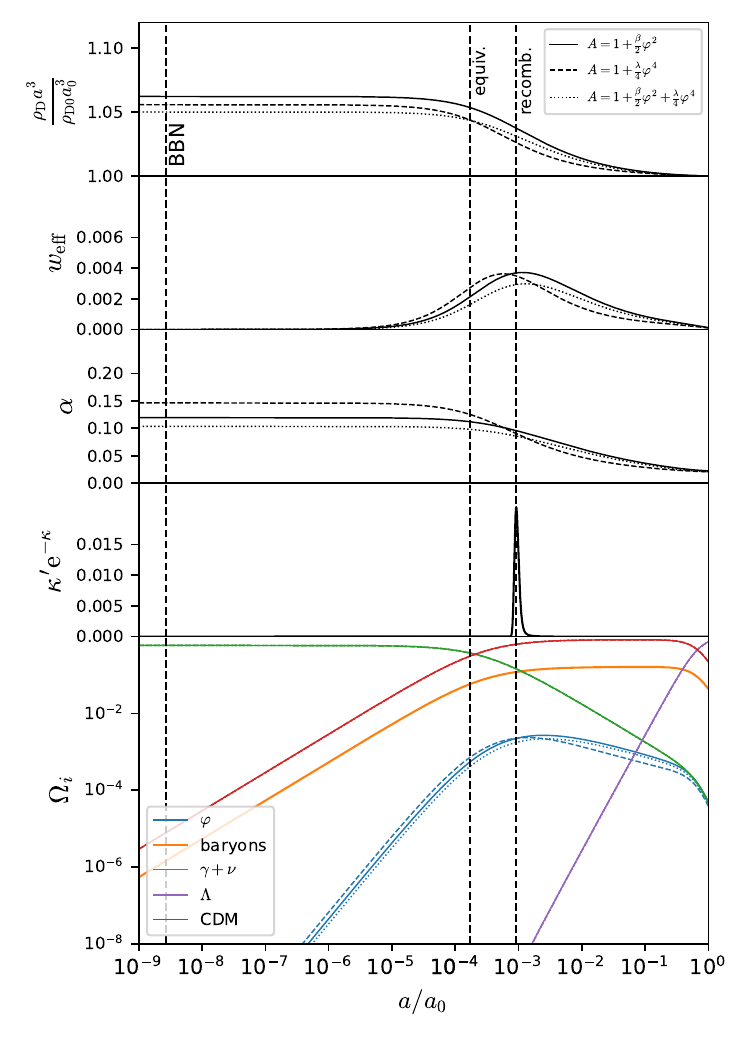}
 	\caption{Background dynamics for the coupling functions~(\ref{e.defmodel1}-\ref{e.defmodelE}) assuming $\xi=1$ for the best fit models whose parameters are given in Table~\ref{tab:2}. From {\it top} to {\it bottom}: evolution of the DM energy density exhibiting the departure from pure dust around recombination; the effective equation of state defined in Eq.~(\ref{e.weff});  the strength $\alpha$ of the scalar interaction compared to the Newton interaction; the visibility function; and the cosmological evolution of the energy densities (in units of $3H^2(a)/8\pi G$ showing that the scalar field energy density remains subdominant at all times. In all cases, initial conditions have been set directly on the radiation era attractor (see \S~\ref{SecScalarFieldDynamics}).}
 	\label{fig:2}
 	\vspace{-0.25cm}
      \end{figure} 

\subsection{Dynamics of DM}\label{SecScalarFieldDynamics}

The background dynamics of DM is affected by the scalar fifth force. Assuming a constant equation of state $w_{\rm D}$, Eq.~(\ref{e.c2}) gives
\begin{equation}
 \rho_{\rm D} \propto a^{-3(1+w_{\rm D})}\left[\frac{A(\varphi)}{A_0}\right]^{4-3(1+w_{\rm D})}.
\end{equation}
Note that when $w_{\rm D}=1/3$, we recover the conformal invariance of Maxwell theory which implies that the radiation does not depend on $A$. We assume that $P_{\rm D}=0$ in this article but provide the equations for any $w_{\rm D}$. We also stress that we have the freedom to assume that only a fraction $\xi$ of the DM energy density today is subjected to the scalar interaction, i.e.
\begin{equation}\label{e.rhoDM}
G \rho_{\rm D} =  G\rho_{\rm D0} a^{-3(1+w_{\rm D})}+  \xi G\rho_{\rm D0} a^{-3(1+w_{\rm D})}\delta_A(\varphi)
\end{equation}
with
\begin{equation}\label{e.54}
\delta_A(\varphi)\equiv \left[\frac{A(\varphi)}{A_0}\right]^{4-3(1+w_{\rm D})} -1.
\end{equation} 
This could be interpreted e.g. as having two DM sectors, one subjected to GR and one subjected to ST gravity and with $(\Omega_{\rm Dv0},\Omega_{\rm Dd0})=(1-\xi,\xi)\Omega_{\rm D0}$. Note that the effect of $\varphi$ on $\rho_{\rm DM}$ can be interpreted as due to  a variation of an effective gravitational constant
\begin{equation}
  G_{\rm eff}(\varphi)=G\left[1+\xi\delta_A(\varphi)\right]
\end{equation}   
since $G\rho_{\rm D}  =G_{\rm eff}\rho_{\rm D0} a^{-3(1+w_{\rm D})}$.  Indeed $G_{\rm eff}$ does not coincide with the gravitational constant~(\ref{e.Gcav}).

Phenomenologically, the fifth force modifies the evolution of the DM energy density. It can be described as an effective equation of state defined as
\begin{equation}
1+ w_{\rm eff} \equiv -\frac{1}{3H}\frac{\dot\rho_{\rm DM}}{\rho_{\rm D}}.
\end{equation}
Using equation~(\ref{e.c2}), it gives
\begin{equation}\label{e.weff}
 w_{\rm eff}(a) =  w_{\rm D}-\frac{1}{3}(1-3w_{\rm D}) \frac{\dd\ln A}{\dd\ln a}.
\end{equation}

To finish, the reduced Friedmann equation takes the form
\begin{eqnarray}
&&E^2(z) = \Omega_{\rm m}(1+z)^{3} + \Omega_{\rm r0}(1+z)^4 + \Omega_{\Lambda0} \nonumber \\
&& \qquad\qquad + \xi\Omega_{\rm D0}(1+z)^3 \delta_A (z)+ \Omega_{\dot\varphi} + \Omega_V.
\end{eqnarray}
The first line corresponds to the standard $\Lambda$CDM while the second gathers all the effects of the scalar interaction. The standard $\Lambda$CDM model is recovered when $\xi=0$ and $w_{\rm D}=0$. Models with $\xi=1$ are called {\it minimal} and have $G_{\rm eff}=GA/A_0$.  Even though we give the full set of background equations for any constants $w_{\rm D}$ and $\xi$ to allow for broader investigations, the following numerical investigations will assume $w_{\rm D}=0$ and $\xi=1$.

The evolution of the background quantities for the three models~(\ref{e.defmodel1}-\ref{e.defmodelE}) are depicted in Fig.~\ref{fig:2}. We can already draw some generic conclusions. As announced, $\varphi$ is always subdominant so that it does not account for dark energy; the energy density of DM or equivalently $G_{\rm eff}$, varies around the last scattering surface, the intensity of the scalar coupling vanishes at low redshift, so that DM enjoys pure Einstein gravity, while it reaches a constant value ranging between $5$ and $7\%$ at high redshifts.

\subsection{Dynamics of the scalar field}

\begin{figure}[htb]
 	\includegraphics[width=\columnwidth]{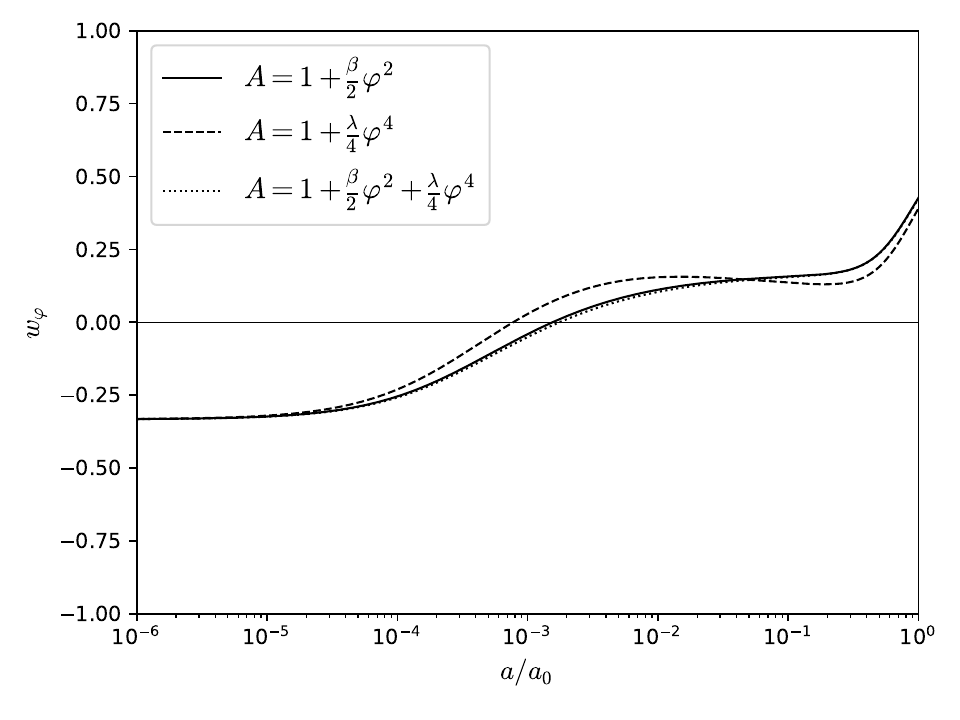}
 	\caption{Equation of state $w_\varphi$ of the scalar field for the best fits of the three models~(\ref{e.defmodel1}-\ref{e.defmodelE}); see Table~\ref{tab:2} for their parameters.}
 	\label{fig:wphi}
 \end{figure}

The intensity of the dark fifth force depends on the evolution of $\varphi$ dictated by the Klein-Gordon equation~(\ref{kgEF}). First, once multiplied by $\dot\varphi$, it gives the conservation equation
\begin{equation}\label{e.cphi}
\dot\rho _\varphi+ 3H(\rho_\varphi + P_\varphi) = - 8\pi G \xi \rho_{\rm D} (1-3w_{\rm D})
\frac{A'}{A}\dot \varphi  
\end{equation}
with the energy and pressure defined in Eq.~(\ref{def.rhophi}).  This equation can be used to define an effective equation of state for the scalar field defined by $\dot\rho _\varphi+ 3H(1+w_\varphi)\rho_\varphi=0$, as
\begin{equation}\label{e.wphi}
w_\varphi = \bar w_\varphi+\frac{8 \pi G\rho_{\rm D}}{3\rho_\varphi}\xi(1-3w_{\rm D})\frac{\dd\ln A}{\dd\ln a}
\end{equation}
with $\bar w_\varphi\equiv P_\varphi/\rho_\varphi$ that depends on $V$; see Eq.~(\ref{def.rhophi}). Indeed in all the models considered in this article for which $V=0$, it reduces to $\bar w_\varphi=1$. Then, it is easy to check that Eqs.~(\ref{e.c2}) and (\ref{e.rhoDM}) imply
\begin{equation}\label{e.ww}
( 8\pi G\xi\rho_{\rm D}  + \rho _\varphi)^. + 3H[8\pi G\xi\rho_{\rm D}(1+w_{\rm D}) + \rho _\varphi (1+ \bar w_\varphi)]=0
\end{equation} 
which is indeed nothing else than Eq.~(\ref{e.DTDM}). It also implies that the DM and $\varphi$ effective equations of state satisfy
\begin{equation}
8\pi\xi G\rho_{\rm D}(w_{\rm eff}-w_{\rm D}) + \rho_\varphi(w_\varphi-\bar w_\varphi)=0,
\end{equation}
which is again obvious from Eqs.~(\ref{e.weff}) and~(\ref{e.wphi}).

Assuming the energy density of the scalar field is initially subdominant, the evolution of the effective equation of state of $\varphi$ for the best fits of three models~(\ref{e.defmodel1}-\ref{e.defmodelE}) is depicted on Fig.~\ref{fig:wphi}. Interestingly, the scalar field behaves as a small curvature term ($w_\varphi\simeq-1/3$) deep in the radiation era while mimicking an almost pressureless fluid around recombination and then increasing toward a stiffer value ($w_\varphi>+1/3$) at low redshift so that it is redshifted away faster than radiation. It is this combination of properties among all the possible properties allowed by these models and that cannot be guessed beforehand that were actually favored by the MCMC to ensure that the effect on the matter power spectrum remains mild and that they evade the problem of a dark radiation component. Furthermore if initially the scalar field dominates the energy budget over radiation, we show below that it is generically attracted toward the solution where it becomes subdominant with $w_\varphi\simeq-1/3$.\\

In order to study the evolution of the scalar field, it is useful to use the time variable $p$ defined by
$$p \equiv \ln a, \qquad a=\hbox{e}^p$$ 
as time variable ranging from $-\infty$ to 0. It  follows that
$$
\dot\varphi=H\varphi',\qquad
\ddot\varphi=H^2\varphi''+ \dot H\varphi'.
$$ 
Inserting these expressions in Eq.~(\ref{kgEF}) and using Eqs.~(\ref{einsteinEF1}-\ref{einsteinEF2}) to express $H^2$ and $\dot H =\ddot a/a-H^2$, the Klein-Gordon equation~(\ref{kgEF}) is rewritten as
\begin{widetext}
\begin{eqnarray}\label{e.KGp}
  &&\frac{2\varphi''}{3-{\varphi'}^2}+\frac{8\pi G\rho(1-w)+2\Lambda+4V}{8\pi G\rho+\Lambda+2V}\varphi' 
   =-\frac{8\pi G\rho_{\rm D}\alpha(\varphi)+2\dd V/\dd\varphi}{8\pi G\rho+\Lambda+2V}
\end{eqnarray}
 with $\rho=\rho_{\rm b}+\rho_{\rm r} +\rho_{\rm D}$. A prime denotes a derivative with respect to $p$ on the field and with respect to $\varphi$ on $A$. The positivity of $H^2$ implies that $3-{\varphi'}^2>0$. Now, when $V=0$,  $\rho_\varphi = H^2{\varphi'}^2$ and  we have the master equation
 \begin{eqnarray}\label{e.master1}
\frac{2\varphi''}{3-{\varphi'}^2}+
&&\left[1- \frac{1}{3}  \frac{\Omega_{\rm r0}\hbox{e}^{-p}-3\Omega_{\Lambda0} \hbox{e}^{3p}}{\Omega_{\rm r0}\hbox{e}^{-p} + \Omega_{\rm b0} + \Omega_{\rm D0}\left[(1-\xi) + \xi\frac{A(\varphi)}{A_0}\right]+ \Omega_{\Lambda0} \hbox{e}^{3p}} \right] \varphi'
 = \nonumber\\
&&\qquad\qquad\qquad\qquad\qquad -\frac{\xi\Omega_{\rm D0} }{\Omega_{\rm r0}\hbox{e}^{-p} + \Omega_{\rm b0} + \Omega_{\rm D0}\left[(1-\xi) + \xi\frac{A(\varphi)}{A_0}\right] + \Omega_{\Lambda0}\hbox{e}^{3p}}
 \frac{\dd}{\dd\varphi}\left(\frac{A(\varphi)}{A_0}\right).
 \end{eqnarray}
 \end{widetext}
We shall now study the 3 usual limiting cases of radiation, matter and cosmological constant dominated eras. To grasp the dynamics of $\varphi$, we assume that $A$ is given by the coupling function~(\ref{e.defmodel1}), that is by a $\Lambda\beta$CDM model, which is actually the limit of all models at small fields.

\subsubsection{Radiation era}

When $\rho_{\rm r}\gg \rho_{\rm b},\rho_{\rm D},\Lambda$ without any assumptions on $\rho_\varphi$.  Eq.~(\ref{e.master1}) simplifies to
 \begin{eqnarray}\label{rdu-phi}
\frac{2\varphi''}{3-{\varphi'}^2}+ \frac{2}{3} \varphi' =  -\frac{\xi\Omega_{\rm D0} }{\Omega_{\rm r0}A_0 } \hbox{e}^{p} A'(\varphi). 
 \end{eqnarray}
We define the time at which  the forcing term on the r.h.s. starts to be non negligible by introducing $p_{\rm eq}$ as 
\begin{equation}
\exp(-p_{\rm eq})\equiv {\Omega_{\rm D0} }/{\Omega_{\rm r0}A_0 }.
\end{equation} 
Interestingly one can estimate, neglecting the baryons and the variation of $A(\varphi)$, that ${\Omega_{\rm D0} }/{\Omega_{\rm r0}A_0 }\sim (1+z_{\rm eq})$, $z_{\rm eq}$ being the redshift at matter-radiation equivalence, so that the driving force on the r.h.s. of Eq.~(\ref{rdu-phi}) is of order
$$
\xi\beta\frac{1+z_{\rm eq}}{1+z}\varphi = \xi\beta\hbox{e}^{p-p_{\rm eq}} \varphi.
$$
It implies, unless $\beta$ is extremely large, that the effect of the non-minimal coupling starts to have a dynamical effect around the equivalence.

First, deep in the radiation era and neglecting the driving force, i.e. for $p\ll p_{\rm eq}$, the field equation 
$$
\frac{2\varphi''}{3-{\varphi'}^2}+ \frac{2}{3} \varphi' = 0
$$
has the general solution for $\varphi(p_i)=\varphi$ and $\varphi'(p_i)=\varphi'_i$,
\begin{eqnarray}
\varphi &=& \varphi_i + \sqrt{3}\left\lbrace {\rm ArgTanh}\left[\frac{\varphi_i'}{\sqrt{3}}\right] \right.    \\
&&\qquad\left.- {\rm ArgTanh}\left[ \frac{\varphi_i'}{\sqrt{\varphi_i^{\prime2}+(3-\varphi_i^{\prime2})\hbox{e}^{2(p-p_i)}}} \right]
\right\rbrace\,,
\nonumber
\end{eqnarray}
and thus
\begin{eqnarray}
\varphi' &=& \frac{\sqrt{3} \varphi_i'}{\sqrt{\varphi_i^{\prime2}+(3-\varphi_i^{\prime2})\hbox{e}^{2(p-p_i)}}}\,.
\nonumber
\end{eqnarray}
One can also conclude that the field asymptotically settles to a constant
$$
\varphi \rightarrow \tilde\varphi_i= \varphi_i + \sqrt{3} {\rm ArgTanh}\left[\frac{\varphi_i'}{\sqrt{3}}\right].
$$
It is easily seen that $\varphi'$ becomes negligible compared to $\sqrt{3}$, so that even if it were dominated initially, it becomes subdominant, $\rho_\varphi\ll\rho_{\rm r}$, rapidly deep in the radiation era. In that regime the field converges toward the solution of $\varphi''+ \varphi' = 0$, that is 
\begin{eqnarray}\label{e.rnr0}
\varphi &=& \varphi_i -\varphi_i'\left[1-\hbox{e}^{p_i-p}\right].
\end{eqnarray}
As a consequence, even if the scalar field dominates at early time, it becomes rapidly subdominant.  Obviously the solution behaves as $\varphi'\propto a^{-1}$. During the radiation era, the Friedmann equation~(\ref{einsteinEF1}) gives $a\propto t^{1/2}$ so that as long as the coupling is negligible in the Klein-Gordon equation~(\ref{kgEF}), $\dot\varphi\propto a^{-3}\sim t^{-3/2}$, i.e. $\varphi\propto t^{-1/2}\propto a^{-1}$. In that regime $\rho_\varphi=H^2\varphi^{\prime2}=H^2\varphi_{\rm init}^{\prime2}a^2_{\rm init}/{a^2}$, hence
$$
\rho_\varphi = 8\pi G\rho_{\rm r} \varphi_{\rm init}^{\prime2} \left(\frac{a_{\rm i}}{a}\right)^2\propto a^{-6}
$$
as expected for a massless scalar field.

However, even in the radiation era, one shall take into account the effect of the driving force that acts before $p_{\rm eq}$ to drive the field toward 0.
The field being in slow-roll,\begin{eqnarray}\label{eq:defK0}
\varphi''+ \varphi' =  -K_0 \hbox{e}^{p-p_{\rm eq}} \varphi,
\end{eqnarray}
where the parameter controlling the onset of the transition is defined in general on the attractor by
\begin{equation}\label{eq:defK0b}
K_0 \equiv \frac{3}{2}\xi \frac{A'(\varphi_i)}{\varphi_i}\,, 
\end{equation}
and since we restrict the current analytic analysis to the $\Lambda\beta$CDM model, it reduces to the constant
\begin{equation}\label{eq:defK0c}
K_0= \frac{3}{2}\xi\beta.
\end{equation}
Consequently, Eq. \eqref{eq:defK0}  is solved in terms of Bessel functions as
\begin{equation}
\varphi(a) = \sqrt{\left(\frac{a_{\rm eq}}{a}\right)}\left[k_1 J_1\left(2\sqrt{K_0\frac{a}{a_{\rm eq}}}\right) + k_2  Y_1\left(2\sqrt{K_0 \frac{a}{a_{\rm eq}}}\right)\right].\nonumber
\end{equation}
As long as $a\ll a_{\rm eq}/K_0$, we recover~(\ref{e.rnr0}), by expanding the Bessel function around $0$, since
\begin{equation}\label{eq:varphia}
\varphi(a) = -\frac{k_2}{\pi\sqrt{K_0}}\frac{a_{\rm eq}}{a}  +\sqrt{K_0}\left[ \frac{2\gamma_E-1}{\pi}k_2+k_1
\right]
\end{equation}
to which we get the slow-roll correction terms
\begin{eqnarray}
&&\frac{\sqrt{K_0}}{\pi} k_2\ln\left(K_0 \frac{a_{\rm eq}}{a} \right)\nonumber\\
&&+
\left[\frac{4\gamma_E-5}{2}k_2+k_1\pi + k_2\ln\left(K_0 \frac{a_{\rm eq}}{a} \right)
\right]\frac{K_0^{3/2}}{2\pi}\frac{a}{a_{\rm eq}}.\nonumber
\end{eqnarray}

\subsubsection{Matter era}

When $\rho_{\rm b},\rho_{\rm D} \gg \rho_{\rm r}, \Lambda$ without any assumptions on $\rho_\varphi$,  Eq.~(\ref{e.master1}) simplifies to
 \begin{eqnarray}
\frac{2\varphi''}{3-{\varphi'}^2}+ \varphi' =  -\frac{\xi\Omega_{\rm D0}\frac{A'(\varphi)}{A_0} }{\Omega_{\rm b0} +\Omega_{\rm D0}\left[(1-\xi) + \xi\frac{A(\varphi)}{A_0}\right] }\,.
\end{eqnarray}
Since $\rho_{\varphi}$ is subdominant at  the end of the radiation era, one can safely neglect ${\varphi'}^2$. For small $\varphi$, 
\begin{eqnarray}
\varphi''+ \frac{3}{2} \varphi' =  -K_0\Xi_0 \varphi,\qquad
\Xi_0= \frac{\Omega_{\rm D0} }{\Omega_{\rm b0}A_0 } \,.
\end{eqnarray}
Hence, the field decays as
$$
\varphi \propto a^{-3/4}\left[ k_1\hbox{e}^{\sqrt{9-16K_0\Xi_0}\ln a}  + k_2\hbox{e}^{-\sqrt{9-16K_0\Xi_0}\ln a}  \right].
$$
with or without oscillations depending on the parameters $K_0$ (theory) and $\Xi_0$ (cosmology). Since $\Xi_0 \sim 10-15$ (see Table~\ref{tab:2}), the condition for oscillations $9-16K_0\Xi_0<0$ teaches us that it requires $\beta\xi\gtrsim 0.02$.

\subsubsection{$\Lambda$ era}

When $\Lambda\gg \rho_{\rm r}, \rho_{\rm b},\rho_{\rm D}$, Eq.~(\ref{e.master1}) simplifies to
 \begin{eqnarray}
\frac{2\varphi''}{3-{\varphi'}^2}+  2\varphi' =  -\frac{\Omega_{\rm D0} \xi }{\Omega_{\rm \Lambda0}A_0 } \hbox{e}^{-3p} A'(\varphi).
\end{eqnarray}
As previously, since $\varphi$ is subdominant, it reduces to
\begin{eqnarray}
\varphi''+ 3 \varphi' =  -K_0 \hbox{e}^{-3(p-p_\Lambda)}\varphi,\quad
\hbox{e}^{3p_\Lambda}\equiv \frac{\Omega_{\rm D0} }{\Omega_{\Lambda0}A_0 },
\end{eqnarray}
the solution of which is
\begin{eqnarray}
\varphi(a) &=& \left(\frac{a}{a_\Lambda}\right)^{-3/2}\left\lbrace k_1 J_1\left[\frac{2}{3}\sqrt{K_0} \left(\frac{a}{a_\Lambda}\right)^{-3/2}\right] \right. \nonumber \\
&&\qquad\qquad\left. + k_2  Y_1\left[\frac{2}{3}\sqrt{K_0} \left(\frac{a}{a_\Lambda}\right)^{-3/2}\right]\right\rbrace.\label{eq:varphiaJY}
\end{eqnarray}
At large $a/a_{\Lambda}$, it behaves as
$$
\varphi(a)\sim \frac{k_1\sqrt{K_0}}{3}  \left(\frac{a}{a_\Lambda}\right)^{-3} - \frac{3k_2}{\pi\sqrt{K_0}}.
$$

\subsubsection{Numerical comparison}

\begin{figure}[htb]
 	\centering
 	\includegraphics[width=0.5\textwidth]{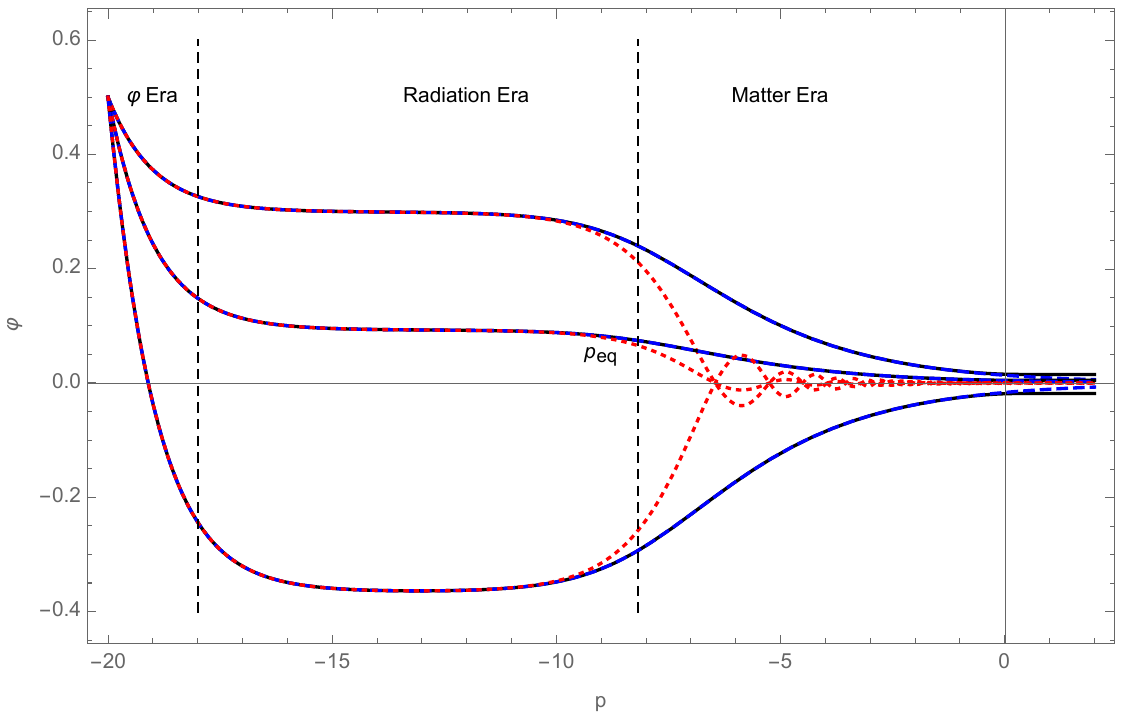} \includegraphics[width=0.5\textwidth]{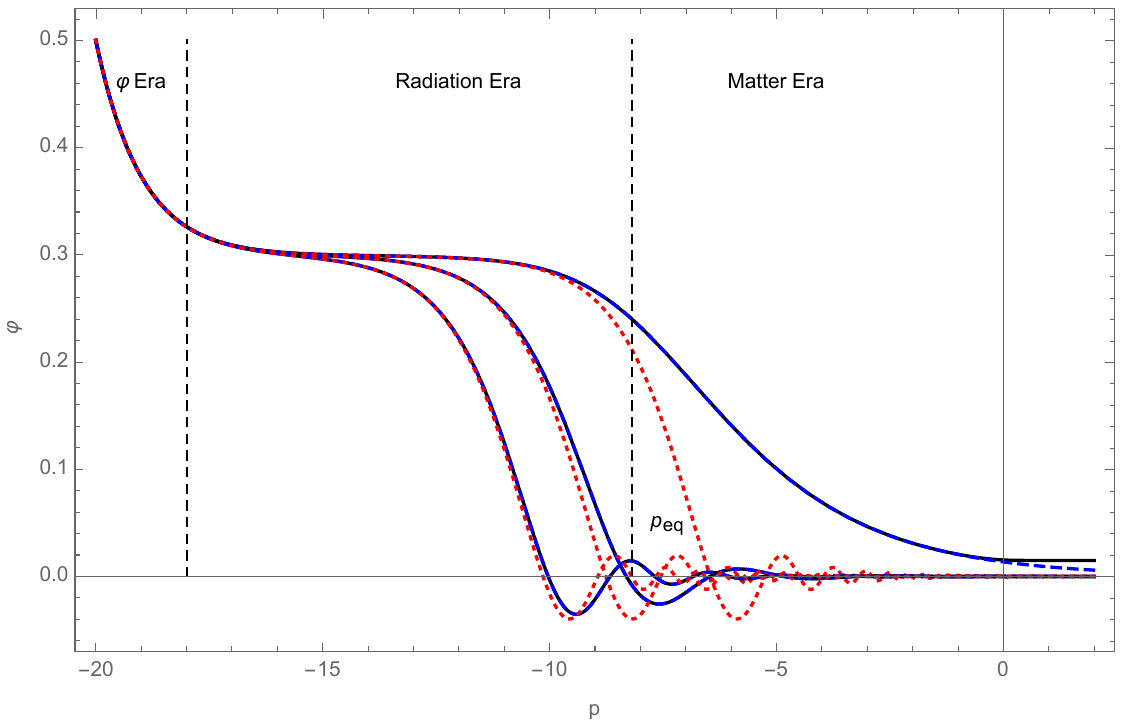}
 	\caption{Evolution of $\varphi$ from a full numerical integration (Solid Blue line) compared to the analytic expressions during the radiation era (Dotted Red line) and matter era (Dashed Black line) assuming a coupling function~(\ref{e.defmodel1}). [{\it Top}] $\beta=0.5$ and $\varphi_i=0.5$ while we vary $\varphi_i'=(-0.2,-0.4,-0.8)$ from top to bottom. The value of $\varphi_i'$ affects the duration of the $\varphi$-dominated era before the field enters the slow-roll radiation dominated attractor. Hence one can set the initial conditions on this attractor deep in the radiation era. [{\it Bottom}] Keeping the initial conditions fixed to $(\varphi_i,\varphi_i')=(0.5,-0.2)$, $\beta$ is varied as $(0.5,5,20)$ from upper to lower curves. }
 	\label{fig:kgrdu}
 	\vspace{-0.25cm}
 \end{figure}

The numerical integration of the full Klein-Gordon equation~(\ref{e.master1}) is compared to these analytical solutions on Fig.~\ref{fig:kgrdu} for the class of models~(\ref{e.defmodel1}). Deep in the radiation era, $\varphi$ is rapidly attracted to a slow-roll attractor and the initial conditions in an eventual $\varphi$-dominated era are erased since its energy density redshifts as $a^{-6}$. Such an early kinetic phase does not imprint cosmological observables but can actually leave a signature on  the primordial  gravitational wave background spectrum (see e.g. Ref.~\cite{Riazuelo:2000fc} for an example). 

Then, during the radiation era, the driving force is controlled by $\beta$ and $p_{\rm eq}$ which explains why it starts to act before equivalence and the earlier for larger $\beta$. In the following matter era, since $\Xi_0\sim 10-15$, $\varphi$ will undergo oscillations  as soon as $\xi\beta\gtrsim 0.02$. The period of the oscillations in $\ln a$ increases with $\beta$. According the the value of the plateau during the radiation era, the field may or may not relax and oscillate around 0. As we can see from Fig~\ref{fig:wphi}, the best-fit solutions does not reach such a behavior but this explains why the effective equation of state of the scalar field is driven to values larger than $1/3$ during the matter era. So the onset of the attraction toward $\varphi=0$ depends on $\beta$ and the time to reach this minimum depends on both $\beta$ and $\varphi_i$. The cosmological constant only induces a small kick in the last $e$-fold.

\subsection{Effective models and reconstruction problem}

At the background level, our model reduces to a form of interaction between DM and a scalar field since the the dark sector is dictated by the set~(\ref{e.c2}-\ref{e.cphi}). It can be rewritten either as a set of interacting fluids as
\begin{eqnarray}\label{e.cpl1}
\left\lbrace 
\begin{array}{l}
\dot\rho_{\rm D} + 3H \rho_{\rm D} (1+w_{\rm D}) = Q, \\
\dot\rho _\varphi+ 3H\rho_\varphi (1+ \bar w_\varphi) =-8\pi G\xi Q
\end{array}
\right.
\end{eqnarray}
with
\begin{eqnarray}\label{e.defQ}
Q = \alpha (\varphi)\rho_{\rm D}(1 -3w_{\rm D})\dot \varphi
\end{eqnarray}
or as two independent components
\begin{eqnarray}\label{e.cpl2}
\left\lbrace 
\begin{array}{l}
\dot\rho_{\rm D} + 3H \rho_{\rm D} (1+w_{\rm eff}) = 0, \\
\dot\rho _\varphi+ 3H\rho_\varphi (1+ w_\varphi) = 0 
\end{array}
\right.
\end{eqnarray}
where the two effective equation of states $w_{\rm eff}$ and $w_\varphi$ are time dependent and related by Eq.~(\ref{e.ww}). This means we have two identical formulation {\em at the background level only}, the first relying on $\lbrace w_{\rm D}, V(\varphi), A(\varphi)\rbrace$ and the second on $\lbrace w_{\rm eff}, w_\varphi\rbrace$. 

\begin{figure}[tb]
 	\centering
 	\includegraphics[width=\columnwidth]{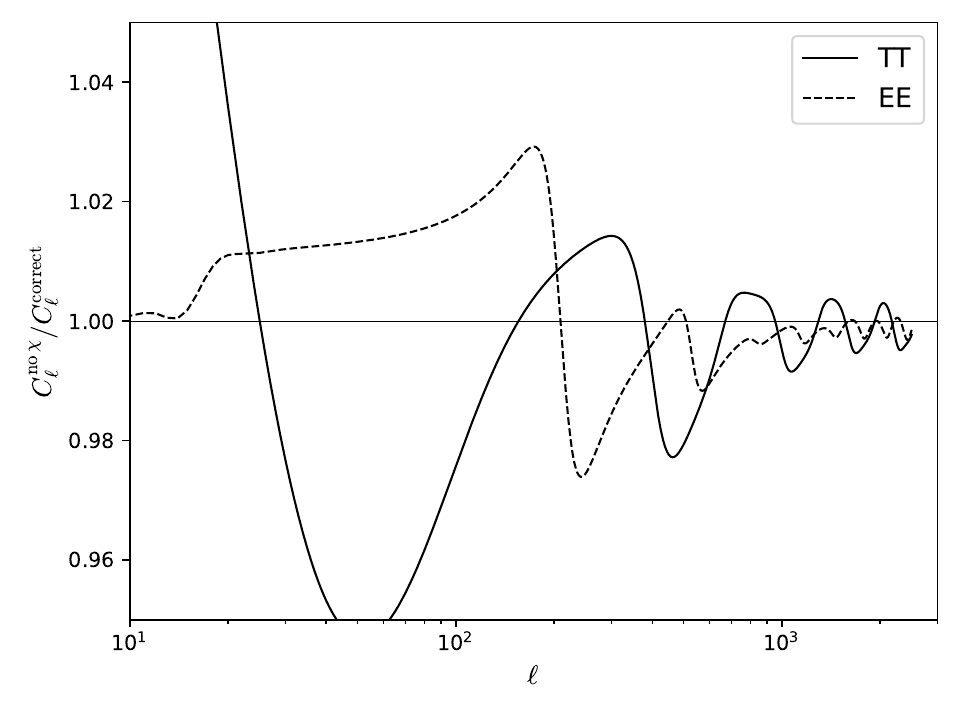}
 	\caption{Example of the impact of a non-consistent treatment of the perturbation on CMD spectra. We compare our best fit model (with consistent background and perturbations) computation to a model with the same background dynamics but which does not take into account the scalar field perturbations. The relative ratio of the corresponding spectra for the best fit of the $\Lambda\beta$CDM model can reach 4\% for the TT spectrum (Solid line) and 2\% for the EE spectrum (Dashed line). This should be compared to the residuals of the Planck data in Fig.~\ref{fig:residuals}. }
 	\label{fig:Clwrong}
 	\vspace{-0.25cm}
 \end{figure}

As discussed in the introduction, many phenomenological models of interacting DM with DE or even some fields of the SM sector have been considered. They are formulated by the choice of equation of states for DE and DM and an interaction terms. For instance,  the interaction term $Q$ was assumed to behave as  $kH\rho_{\rm DE}$~\cite{DiValentino:2019jae},  $kH\rho_{\rm D}$~\cite{Wang:2004cp,Kumar:2016zpg} which means that $\rho_{\rm D}\propto a^{-3+k}$ that corresponds to a model with $w_{\rm D}\not=0$ but without interaction with $\varphi$, $3k(1+w_{\rm DE}) H\rho_{\rm D}$ or  $3k(1+w_{\rm DE}) H(\rho_{\rm D}+\rho_{\rm DE})$~\cite{Yang:2018euj} and $3k(1+w_{\rm DE}(a)) H\rho_{\rm DE}$~\cite{Pan:2019gop}. First, the comparison of these quantities with the interaction term~(\ref{e.defQ}) in our theory shows some drastic differences: (1) all these models are analytic in $\rho_{\rm DE}$ which means while (\ref{e.defQ}) is proportional to $\dot\varphi$, that is $\sqrt{\rho_\varphi+P_\varphi}$, and (2) the strength of the interaction $\alpha$ depends on $\varphi$  hence adapting to $H^2$ through the Klein-Gordon equation. These two features cannot be reproduced by direct couplings between the two dark components. Indeed, one can use our insight to express $\varphi$ as some inverse function of $4V=\rho_\varphi-P_\varphi$ so that, in a purely phenomenological way our model could be mimicked, at the background level only, by an interaction term of the form
$$
Q \sim f(\rho_\varphi-P_\varphi)\rho_{\rm D}(1-3w_{\rm D}) \sqrt{\rho_\varphi+P_\varphi}.
$$
Indeed, the interaction term $Q$ entering the continuity equations at the background level is only the projection of a general interaction term $Q^\mu$, i.e. the projection of Eq.~(\ref{FEq_3}) along the preferred 4-velocity of comoving observers (see e.g. Ref.~\cite{Uzan:1998mc} for a generic discussion of interacting species)
$$
u_\nu\nabla_\mu T_{\rm(DM)}^{\mu\nu} =  \alpha(\varphi) T^{\rm(DM)}_{\sigma\rho} g^{\sigma\rho} u_\nu\partial^\nu\varphi \equiv Q.
$$
As already explained, choosing a phenomenological interaction term or DE equation of state does not give access to a full theory needed in particular to derive a consistent perturbation theory, even if it could reproduce the background equations. The consequences of not treating the perturbation in a consistent way with the background is illustrated in Fig.~\ref{fig:Clwrong} in which we have implemented the CMB computation in a model with the same background as our best-fit models but without taking into account $\varphi$ and its coupling at perturbation level. This leads to typical variations of typically of order 2\% which is much larger than the error bars from the Planck data (compare to Fig.~\ref{fig:residuals} to be convinced that the effects are dramatic). This highlights the difficulty to draw robust conclusions on a purely phenomenological model.

It is interesting to stress that the background equations derived in the previous sections can be used to tackle the reconstruction problem, i.e. to identify a theory that would produce a determined dark sector at the background level. One may want to reproduce a given dark sector history, that is impose a solution of the form
$$
\lbrace \hat\rho_{\rm DE}(a), \hat\rho_{\rm DM}(a) \rbrace,
$$
with a hat to recall that this corresponds to arbitrary phenomenological chosen functions. Since we have two free functions, one may try to reconstruct the theory that will contain this solution. For the two components we can derive their equations of state $\hat w_{\rm DE/DM}(a)=-(\dd\ln \hat\rho_{\rm DE/DM}a^3/\dd\ln a)/3$. These equations of state are the one entering the system~(\ref{e.cpl2}), that is they shall coincide with $\lbrace w_{\rm eff}, w_\varphi \rbrace$. This determines the coupling function from Eq.~(\ref{e.rhoDM}) as
\begin{equation}\label{e.rec1}
A(a)\propto \left[\hat \rho_{\rm DM}(a)a^{3(1+w_{\rm D})}\right]^{\frac{1}{4-3(1+w_{\rm D})}},
\end{equation}
$w_{\rm D}$ being free. Then one deduces that
$$
\bar w_\varphi (a)= \hat w_{\rm DE}(a)+\frac{8\pi G\xi \hat \rho_{\rm DM}(a)}{\hat \rho_{\rm DE}(a)} \left[ \hat w_{\rm DM}(a) - \hat w_{\rm DE}(a) \right].
$$
Then, from Eq.~(\ref{def.rhophi}) we deduce that
\begin{equation}\label{e.rec2}
\left\lbrace 
\begin{array}{l}
4V(a) = \hat\rho_{\rm DE}(a) \left[1-\bar w_\varphi (a)\right], \\
\left(\frac{\dd\varphi}{\dd a}\right)^2=\frac{1}{2\hat H^2(a) a^2}\hat\rho_{\rm DE}(a) \left[1+\bar w_\varphi (a)\right]
\end{array}
\right..
\end{equation}
Hence Eqs.~(\ref{e.rec1}) and~(\ref{e.rec2}) provide $\lbrace A(a), V(a), \varphi(a)\rbrace$. One cannot prove the theory will be well-defined but this can serve as a guide to construct an action mimicking a chosen behavior and hence to draw its cosmological signature consistently.

\subsection{Generic features of the models}

Before we proceed to the computations of the cosmological observable, we can already mention the following conclusions for any model with $V=0$ and $A\sim1+\beta\varphi^2/2$ at small field:
\begin{itemize}
\item First, Fig.~\ref{fig:2}  confirms that $\rho_\varphi$ is negligible (at most $0.27\%$ of the total matter content) in the minimal $\Lambda\beta$CDM model. Hence  the new scalar degree of freedom modifies gravity in the DM sector but not the expansion history directly. 
\item Then, the DM fifth force vanishes in the late universe ($\alpha\rightarrow0$) and saturates to $\alpha=0.107$ in the early universe while the effective coupling,  $G_{\rm eff}$,  varies of  $A_\infty/A_0-1\sim 6.4\%$ between constant values long before equivalence and after recombination.
\item Phenomenologically the main variation starts around equivalence and mostly takes place around recombination in an extended redshift region. It happens generically on this scale due to the coupling to DM so that the variation of $\varphi$ in the radiation era is dictated by the parameters $\beta$ and $p_{\rm eq}$.
\item The conservation equations imply that $\rho_{\rm D}$ transfers to $\rho_\varphi$. The latter can enjoy a large variety of dynamics given the parameter $\beta$ and the initial conditions. As soon as $\varphi$ undergoes damped oscillations, $\rho_\varphi$ will scale as $a^{-6}$ so that part of the DM is actually taken out of the matter budget since it transfers to a component that redshifts faster than radiation. However the best-fit models have the property that $\rho_\varphi$ behaves as an almost-dust like fluid around the recombination before getting a stiffer equation of state that eventually becomes larger than $1/3$ at low redshifts. This is a key difference with DM decaying into dark radiation that scales as $a^{-4}$. Hence, and as we shall later see, our models evade the generic problem of the suppression of the matter power spectrum.
\item Both the sound horizon $r_s$ and the angular distance to the CMB are the same as in the standard $\Lambda$CDM. This is a key feature that convinces us these models will reproduce the CMB power spectra of the $\Lambda$CDM to a high accuracy since $\theta_*$ is identical.
\item This already highlights why these models can alleviate the Hubble tension: $\rho_\varphi$ being negligible, one gets the same background sound horizon and distance to the last scattering surface as for a standard $\Lambda$CDM but with a higher $H_0$ and a lower $\Omega_{\rm D0}$, and hence a sizeably larger $\Lambda$. As a consequence of the fact that gravity is stronger at high-$z$, the effective cosmological parameter controlling the CMB is $8\pi G_{\rm eff}\rho_{\rm D_0}/3H_0^2 \sim (A_\infty/A_0)\Omega_{\rm D0}$. 
\end{itemize}
Since these models compensate a higher $H_0$ by a lower $\Omega_{\rm D0}$, they predict a younger universe (typically 13.55~Gy for the best-fit of the minimal $\Lambda\beta$CDM model instead of 13.79~Gy for the $\Lambda$CDM, as we shall see in details once confronted with data).

\begin{figure}[htb]
 	\centering
 	\includegraphics[width=0.95\columnwidth]{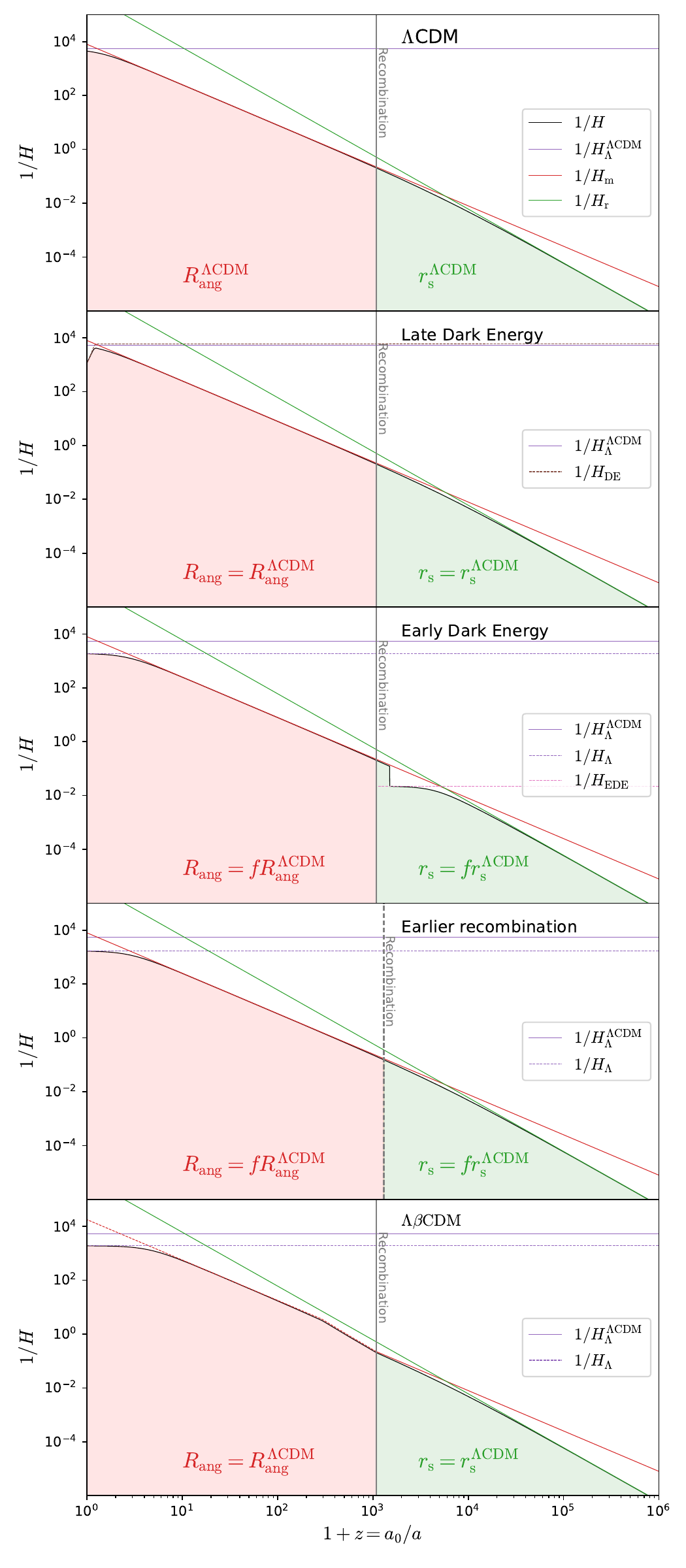}
 	\caption{Qualitative comparison of the existing families of solutions to the Hubble tension with exaggerated and idealized modifications (see text for explanations). Modifications with respect to the $\Lambda$CDM are drawn in dashed lines.}
 	\label{fig:Solutions}
 	\vspace{-0.25cm}
 \end{figure}

Fig.~\ref{fig:Solutions} compares our new models to the three existing categories of models (namely late and early dark energy models and earlier recombination models) to highlight why they differ from the usual solutions investigated so far. In a spatially Euclidean universe, $r_s$ and $R_{\rm ang}$ are both defined as integrals of $1/H(z)$, the former before recombination, and the latter after recombination (see introduction). This is the reason why we plot $1/H(z)$ so that the area below the curve in their respective redshift range provides a visualization of $r_s$ and $R_{\rm ang}$. Note that since we use logarithmic scales for visualisation purposes, these areas are meant as the ones of the corresponding plot with linear scales. We also include the $1/H_i$ for the various components of the Universe ($i$=matter, radiation, $\Lambda$) with $H_i(z)=\sqrt{8\pi G\rho_i(z)/3}$ the Hubble factor that would be obtained from the Friedmann equation if all other species density are set to zero (hence $1/H_i> 1/H$ since $H^2=\sum H_i^2$). For each class of solutions, the angular size of the acoustic peak constrains the ratio of $r_s/R_{\rm ang}$ to be identical to its $\Lambda$CDM value even though the cosmological constant is changed so as to lead to a larger Hubble rate today. For the early dark energy and earlier recombination models, this is achieved by reducing both lengths by a factor $f<1$. This figure clarifies the different roads taken to solve the Hubble tension and highlights the novelty of our class of models.

\section{Large scale structures}\label{sec3}

To compare our model to CMB+BAO data and the Hubble diagram, we need to implement it in a Boltzmann code. To this end, we need the linear cosmological perturbation equations. We can choose to define the perturbations in any gauge which is fully fixed since observables, e.g. CMB anisotropies, must be gauge-invariant. The most popular gauges for numerical integrations are the Newtonian gauge and the dark matter comoving synchronous gauge. In Newtonian gauge and conformal time $\eta$, the spacetime metric for the scalar mode is
 \begin{equation}\label{e.metpert}
 \dd s^2 = 	a^2(\eta)\left[-(1+2\Phi)\dd \eta^2 + (1-2\Psi)\gamma_{ij}\dd x^i \dd x^j\right],
\end{equation}
whereas in synchronous gauge it is decomposed with the $h$ and the $E$ potentials, as 
\begin{eqnarray}\label{e.metper2}
  &&\dd s^2 = 	a^2(\eta)\left[-\dd \eta^2 + \left(1+\frac{h}{3}\right)\gamma_{ij} \dd x^i \dd x^j \right.\nonumber\\
  &&\qquad\qquad\left.-  (\partial_i \partial_j - \frac{1}{3}\Delta \gamma_{ij}) E \dd x^i \dd x^j \right].
\end{eqnarray}

In the next section we collect the perturbation equations in the Newtonian gauge which are physically more transparent since the potentials $\Phi$ and $\Psi$ match the usual Newton potential for small scales. In the subsequent section we detail the numerical implementation and we explain why we must eventually resort to the synchronous gauge for purely numerical reasons, and provide the corresponding equations. Hereafter, a prime denotes a derivative with respect to $\eta$. 

\subsection{Perturbation equations}

The perturbation equations for baryons and radiation are unchanged but we get extra-terms in the Einstein equations and in the conservation equation for DM and $\varphi$. We decompose the scalar field as $\varphi=\varphi(t)+\chi$ in Newtonian gauge.

First, the Einstein equations~(\ref{FEq_1}) become
\begin{eqnarray}
&&\Phi-\Psi =  8\pi G a^2 P_{\rm r}\pi_{\rm r} \label{e.pert1}\\
&&\Psi'+3{\cal H}\Phi=-4\pi G a^2\rho(1+w)v + \varphi'\chi   \label{e.pert2} \\
&&(\Delta+3K)\Psi-3{\cal H}(\Psi'+{\cal H}\Phi) =4\pi G a^2\rho \delta \nonumber\\
&& \qquad\qquad+\varphi'\chi'-{\varphi'}^2\Phi+a^2\frac{\dd V}{\dd\varphi}\chi  \label{e.pert3}
\end{eqnarray}
where e.g. $\rho(1+w)v$ includes all the components, $\rho(1+w)v=\sum_i\rho_i(1+w_i)v_i$ with $i={\rm r,b,D}$ and $\pi_{\rm r}$ is the anisotropic stress of the radiation component (including both photon and neutrinos).

For the component of the visible sector (b, r and possibly a part of the DM not coupled to $\varphi$), we have the standard equations~\cite{Peter:2013avv,Mukhanov:2005sc}
\begin{eqnarray}
&& \delta'+3{\cal H}(c_s^2-w)\delta= -(1+w)(\Delta v- 3\Psi')   \label{e.pert4} \\
&& v'+{\cal H}(1-3c_s^2)v = -\Phi-\frac{c_s^2}{1+w}\delta.  \label{e.pert5}
\end{eqnarray}
Then, for DM, the interaction term is given  by $Q^\nu=\alpha(\varphi) T^{\rm(DM)}_{\sigma\rho} g^{\sigma\rho} \partial^\nu\varphi$ so that $\nabla_\nu T^\nu_\mu=Q_\mu$ reduces to
\begin{eqnarray}
&& \delta_{\rm D} '= -(\Delta v_{\rm D} - 3\Psi') + \left[\alpha\chi' +\newbeta \varphi'\chi \right]  \label{pertDM1} \\
&& v_{\rm D} '+{\cal H}v_{\rm D} = -\Phi  - \alpha \varphi' v_{\rm D} -\alpha\chi \label{pertDM2}
\end{eqnarray}
and one can note that
\begin{equation}\label{e.trick}
 \alpha\chi' +\newbeta \varphi'\chi = \left(\alpha \chi\right)'.
\end{equation}
To finish,  for the scalar field we get
\begin{eqnarray}\label{pertKG}
&& \chi''+2{\cal H}\chi' - \Delta\chi +a^2\frac{\dd^2 V}{\dd\varphi^2}\chi =\left[2(\varphi''+2{\cal H}\varphi')\Phi \right.  \nonumber\\
&&\quad\quad \left.+ \varphi'(\Phi'+3\Psi')\right]-4\pi G\xi\rho_{\rm D}a^2\left[\alpha \delta_{\rm D} +\newbeta\chi  \right].
\end{eqnarray}

\subsection{Numerical implementation}

\begin{figure}[!htb]
 	\centering
	 \includegraphics[width=0.4\textwidth]{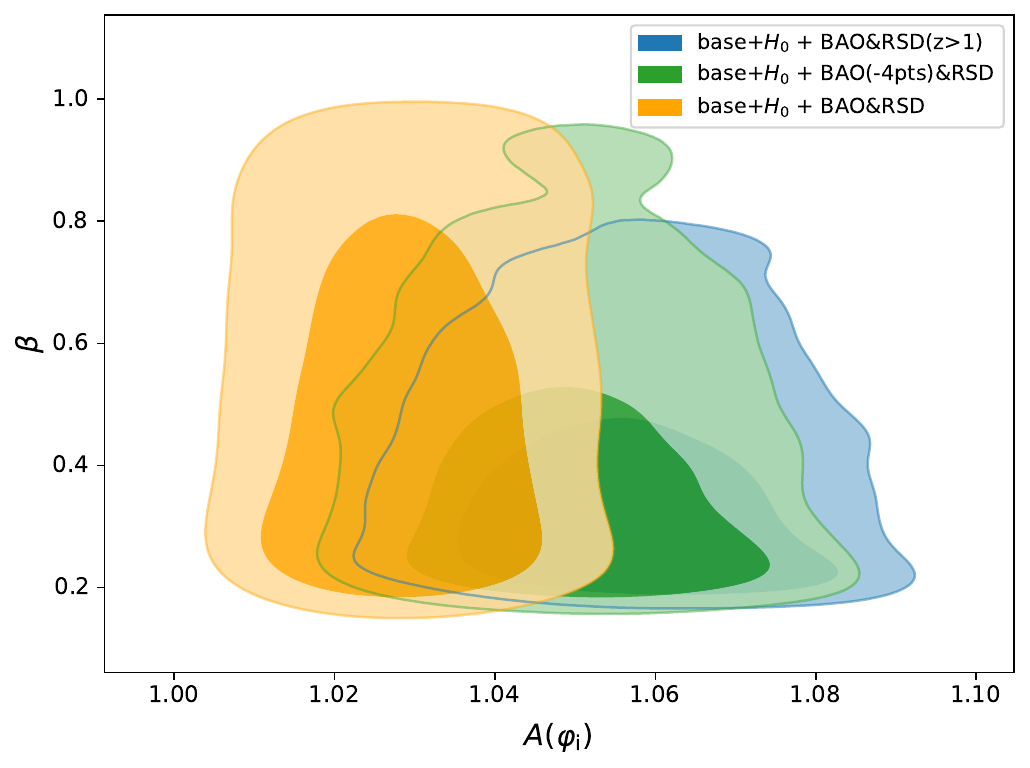}
 	\caption{Marginal distribution of the background scalar field initial conditions of the $\Lambda\beta$CDM model. $\beta$, being related to $K_0$ [see Eq.~\eqref{eq:defK0b}] controls the onset of the transition, whereas $A(\varphi_i)$ dictates the amplitude of energy density extracted from DM by the scalar field. This shows that the best-fit models do not require a fine tuning of their initial conditions.}
 	\label{fig:BetaPhi}
 	\vspace{-0.25cm}
\end{figure}

The model is implemented both at the background and perturbations levels as a modification (available upon request) of the CLASS code \cite{CLASS}. A shooting method is used to ensure the desired final energy density of cold dark matter, given that it no longer scales as $1/a^3$. Also the total energy budget is adjusted with another shooting method since there is a residual energy density in the scalar field not known a priori. The marginal distribution in the $\Lambda\beta$CDM model is shown in Fig.~\ref{fig:BetaPhi}. While the initial conditions need to be adjusted for the scalar field to provide a viable solution, it is important to stress that (1) the space of solutions contains trajectories that provide a better fit than the $\Lambda$CDM and (2) that there is no fine tuning on the parameters $(\varphi_i,A_i)$.

Numerically, the background scalar field is set directly onto its radiation era attractor, that is assuming $k_2=0$ in Eq.~\eqref{eq:varphia}. The derivative of the field is related to the field initial value thanks to the lowest order expansion of Eq.~\eqref{eq:varphiaJY} which then yields $\varphi(a) = \varphi_i[1 - K_0a/(2 a_{\rm eq})] + {\cal O}(a^2)$. The expression of $K_0$ for the $\Lambda\beta$CDM model is given by Eq.~\eqref{eq:defK0c}, and for other models we use the general form~\eqref{eq:defK0c} which, except for the model~(\ref{e.defmodel1}), depends on $\beta$ and $\varphi_i$.

The numerical implementation in Newtonian gauge is not straightforward since the perturbed Klein-Gordon equation \eqref{pertKG} requires the knowledge of both $\Phi'$ and $\Psi'$, but only the latter is given by the Einstein equation \eqref{e.pert2}, whereas the equation \eqref{e.pert1} can only be used to obtain $\Phi$, and getting the derivative of this equation can be numerically unstable. We rather choose to work in synchronous gauge~\cite{Ma:1995ey,Archidiacono:2022iuu} but we have verified that the results are extremely similar to Newtonian gauge results when $\Phi'$ is wrongly replaced by $\Psi'$ in the Klein-Gordon equation.

To that goal, first we need Eqs.~(\ref{pertDM1}-\ref{pertDM2}) and~(\ref{pertKG}) in synchronous gauge,
\begin{eqnarray}
&& \delta_{\rm D} '= -\Delta v_{\rm D}  - \frac{1}{2}h' + \left[\alpha \chi' + \newbeta \varphi'\chi \right]   \\
&& v_{\rm D} '+{\cal H}v_{\rm D}  = \alpha (\chi + \varphi' v_{\rm D} )
\end{eqnarray}
while for the scalar field it is simply
\begin{eqnarray}
&& \chi''+2{\cal H}\chi' - \Delta\chi +a^2\frac{\dd^2 V}{\dd\varphi^2}\chi =-\frac{1}{2}h'\varphi' 
\nonumber\\
&&\qquad\qquad\qquad -4\pi G\xi\rho_{\rm D} a^2 \left[ \alpha \delta_{\rm D} +\newbeta\chi \right].
\end{eqnarray}
The dynamical equation dictating the evolution of $h$, obtained from the perturbed Einstein equations, can be found in Ref.~\cite{Ma:1995ey}.
We set the initial conditions $\chi = -(\alpha/\newbeta) \delta_{\rm D}$, but $\chi'=0$ since it then reaches rapidly its slow-rolling attractor. We consider the usual adiabatic conditions for the other species.
In addition we set the initial condition $v_{\rm D}=0$, which completely fixes the synchronous gauge. Note however that there is a key difference with the usual $\Lambda$CDM-comoving synchronous gauge: the scalar force in the DM sector implies that $v_{\rm D}$ does not vanish at all times since DM particles are not following geodesic of the metric $g_{\mu\nu}$ and the cold dark matter Euler equation must be implemented consistently.

\subsection{Mode evolution}

Let us illustrate some differences of the evolution of our models compared to the standard $\Lambda$CDM.

\begin{figure}[htb]
 	\centering
	 \includegraphics[width=0.4\textwidth]{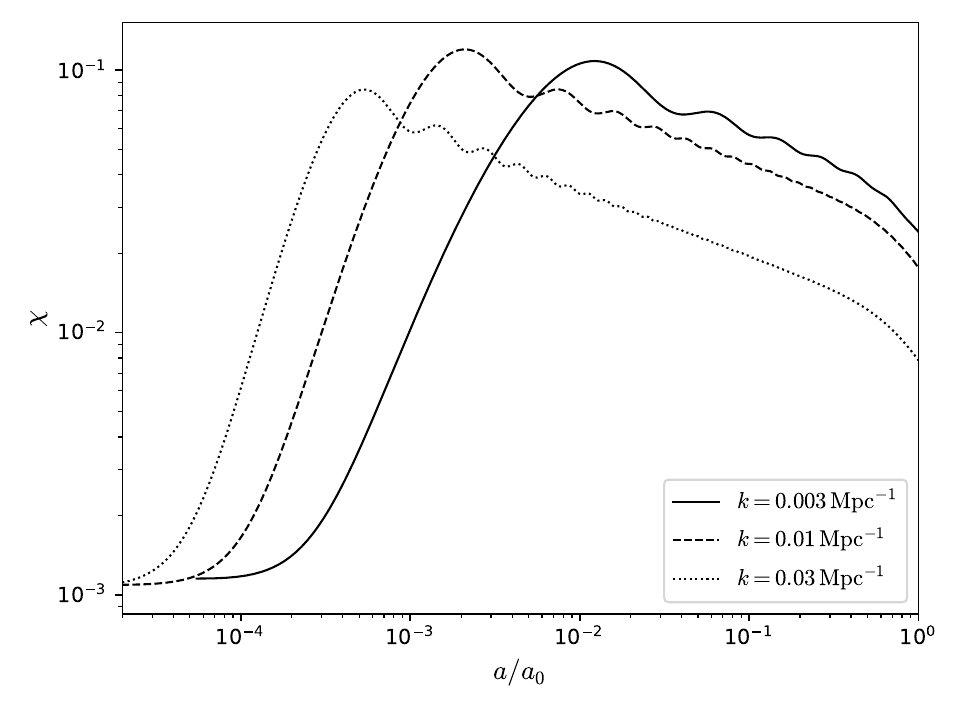}\\
 	\includegraphics[width=0.4\textwidth]{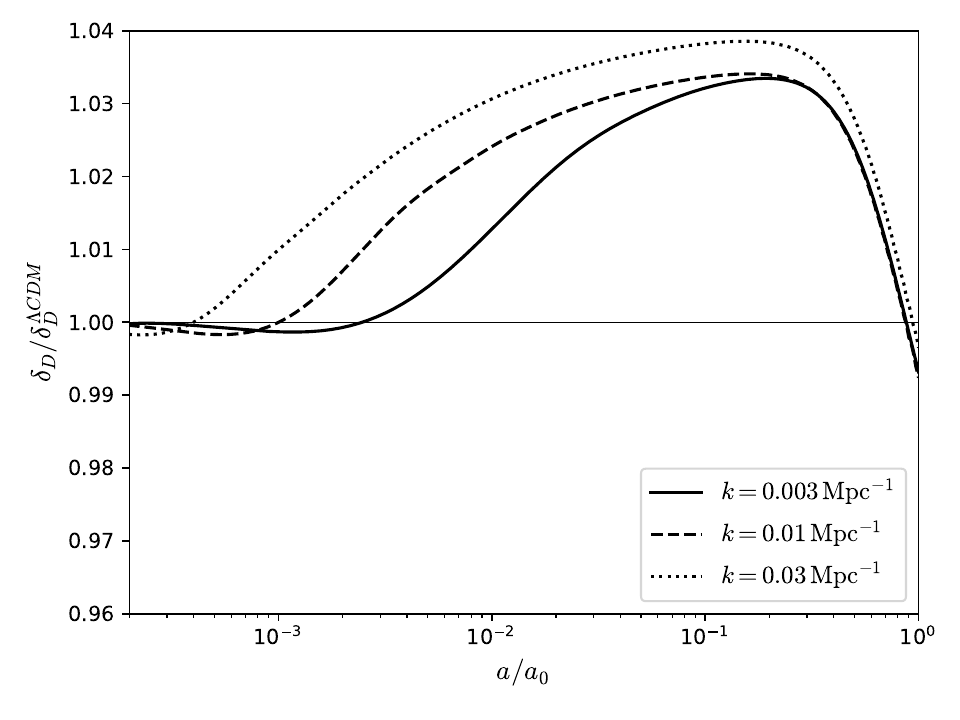}\\
        \includegraphics[width=0.4\textwidth]{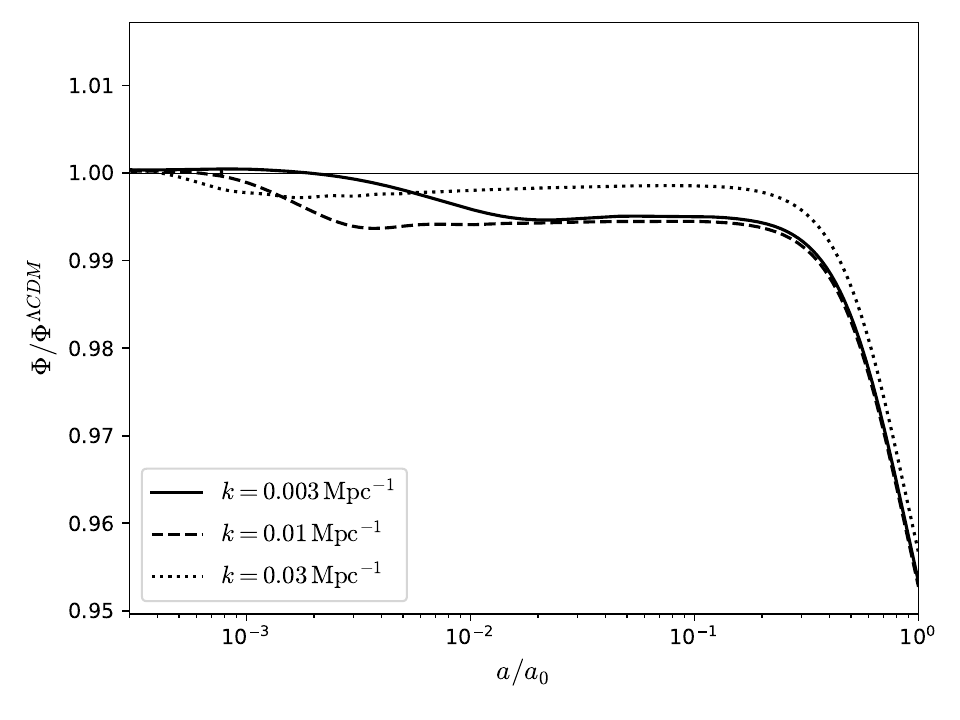}
 	\caption{Evolution of perturbations. [{\it Top}] The scalar field perturbations $\chi$ grow between equivalence and recombination before decaying slowly.
	[{\it Middle}] Relative evolution of $\delta_{\rm D}$ with respect to the $\Lambda$CDM best fit. [{\it Down}] Relative evolution of the gravitational potential $\Phi$. Overall differences with respect to $\Lambda$CDM structure formation are small, hence the amplitude of structure, $\sigma_8$ is expected to be only marginally modified.}
 	\label{fig:Modes}
 	\vspace{-0.25cm}
\end{figure}

To compare the evolution of DM and baryon perturbations, let us combine Eqs.~(\ref{pertDM1}-\ref{pertDM2}) as
\begin{eqnarray}
&& \left( \delta_{\rm D} -\alpha\chi-3\Psi\right)'= - \Delta v_{\rm D}   \label{pertDM1b} \\
&& v_{\rm D} '+\left({\cal H}+ \alpha \varphi' \right)v_{\rm D} = -\Phi  -\alpha\chi  \label{pertDM2b}.
\end{eqnarray}
This is no surprise from the analysis of appendix~\ref{appB} since the first equation is the standard conservation equation for the density contrast $\tilde\delta_{\rm D}=\delta_{\rm D} -4\alpha\chi$ in the gravitational field defined by the two potentials $\tilde\Psi=\Psi-\alpha\chi$ and $\tilde\Phi=\Phi+\alpha\chi$; see Eqs.~(\ref{e.JF2}) and~(\ref{e.JF3}). This explicitly highlights the contribution from the scalar field. Note however that $\tilde\delta_{\rm D}$ and $\tilde v_{\rm D}=v_{\rm D}$ are sourcing the Einstein equations for $\Phi$ and $\Psi$. They combine to give
\begin{eqnarray}
&& \left[\tilde a \left(\tilde \delta_{\rm D} -3\tilde \Psi\right)'\right]'= \tilde a \Delta\tilde\Phi, \nonumber
\end{eqnarray}
that is in an equation for the density contrast in the flat slicing gauge of the DM-Einstein metric. Indeed, the DM is minimally coupled to the metric $\tilde g_{\mu\nu}$ so that cosmologically it experiences the scale factor $\tilde a= aA$; see Eq.~(\ref{e.JF2}). This implies that $\tilde\rho \tilde a^3 = \rho a^3/A$ remains constant. At the perturbation level, the relevant quantity is the $\tilde g$ metric flat slicing gauge invariant perturbation density contrast.

This equation for DM can be combined with the one for the baryons after recombination by defining the relative entropy and velocity as
\begin{equation}
 S\equiv \frac{\delta_{\rm D}}{1+w_{\rm D}} - \frac{\delta}{1+w}, \qquad {\cal V}\equiv v_{\rm D}-v.
\end{equation}
From Eqs.~(\ref{e.pert4}-\ref{e.pert5}) and~(\ref{pertDM1}-\ref{pertDM2}), we get
\begin{equation}
S'+\Delta V =    \left(\alpha\chi\right)',
\quad
V'+{\cal H}V = - \alpha \varphi' v_{\rm D}-\alpha\chi
\end{equation}
leading to
\begin{equation}
\left[a\left(S-\alpha\chi\right)'\right]' =a \alpha \Delta \left[\varphi' v_{\rm D}+\chi\right].
\end{equation}
Indeed baryons are coupled to photons so they also feel a non-vanishing force at the perturbation level before decoupling. This effect of entropy generation will also occur in all models with $\xi\not=1$ since then there will be two DM sectors [see definitions below Eq.~(\ref{e.54})], with only one of them subjected to the scalar fifth force, such that models with $\xi\not=1$ generically develop DM-DM isocurvature modes revealing the scalar force. The entropy generation arises from the fact that the DM component coupled to $\varphi$ is comoving in the metric $\tilde g_{\mu\nu}$ so that $v_{\rm D}$ ``feels" the effective Hubble parameter $(aA)^./(aA)$ and the potential $\tilde\Phi$. This is a cosmological manifestation of the violation of the equivalence principle between the visible and dark sectors.

Fig.~\ref{fig:Modes} illustrates the evolution of the scalar field perturbations for the best fit model with coupling~(\ref{e.defmodel1}). This extra-component compared to the standard $\Lambda$CDM starts to grow around the equivalence and earlier for smaller wavelengths. Whatever $k$ they peak to $0.1$ and then slowly decay while oscillating when sub-horizon. Even though it is difficult, and even not really meaningful to compare the evolution of the density contrasts and potentials in two different models because the perturbations are not living on the same background, such  comparison gives a flavor of the effects on the CMB and matter power spectra. First the effect is smaller than 4\% independently of the wave number $k$ so that the amplitude of structures, $\sigma_8$, is expected to be only marginally modified. Second, the potentials witness a shift smaller than 1\% after recombination and remain constant almost until a redshift of order $5$ when they drop by several percents. Hence we shall not expect a sizable difference in the integrated Sachs-Wolfe contribution to the CMB.

\section{Comparison to observations}\label{sec4}

\subsection{Cosmological data}\label{secdata}

In order to test its power on the Hubble tension we perform a MCMC analysis with Cobaya~\cite{Cobaya}. We consider a baseline dataset which consists in
\begin{itemize}
\item CMB data from Planck : low and high $\ell$ temperature and polarization, with CMB lensing \cite{Planck:2019nip};
\item DES Y1 : weak lensing and galaxy correlations (3x2pt) \cite{DES:2017myr};
\item supernovae (SN) data from the Pantheon sample~\cite{Pan-STARRS1:2017jku}.
\end{itemize}

BAO data are then combined to this baseline in different ways. First, we split BAO measurements into a low-$z$ (noted $z<1$ hereafter) and high-$z$ (noted $z>1$ hereafter) datasets. 

The low-$z$ consists in
\begin{itemize}
\item 6dF : a distance measurement at $z_{\rm eff} = 0.106$ \cite{2011MNRAS.416.3017B};
\item SDSS DR7 [main galaxy sample (MGS)] : $R_{\rm ang}$ and $D_H \equiv 1/H$ at $z_{\rm eff} = 0.15$ \cite{Ross:2014qpa};
\item SDSS DR12 : $R_{\rm ang}$, $D_H$ and redshift space distorsions (RSD) at $z_{\rm eff} = 0.38$ and $z_{\rm eff} = 0.50$~\cite{BOSS:2016wmc};
\item luminous red galaxies (LRG) of SDSS DR16 :  $R_{\rm ang}$, $D_H$ and RSD at $z_{\rm eff} = 0.698$ \cite{eBOSS:2020yzd},
\end{itemize}

The high-$z$ dataset consists in
\begin{itemize}
\item emission line galaxies (ELG) of SDSS DR16 : $R_{\rm ang}$, $D_H$ and RSD at $z_{\rm eff} = 0.85$~\cite{eBOSS:2020yzd};
\item quasistellar objects (QSO) of SDSS DR16 : $R_{\rm ang}$, $D_H$ and RSD at $z_{\rm eff} = 1.48$;
\item Lyman-$\alpha$ (Ly$\alpha$) absorption lines of SDSS DR16 : $R_{\rm ang}$, $D_H$ at $z_{\rm eff} = 2.33$;
\item Ly$\alpha$-QSO cross-correlations in SDSS DR16 : $R_{\rm ang}$, $D_H$ at $z_{\rm eff} = 2.33$;
\end{itemize}

Finally the local measurement of $H_0$ from the recent SH0ES results~\cite{Riess:2021jrx} is combined with the baseline dataset and the chosen BAO data.

\subsection{Analysis of the $\Lambda$CDM model}

\begin{figure}[htb]
 	\centering
 	\includegraphics[width=\columnwidth]{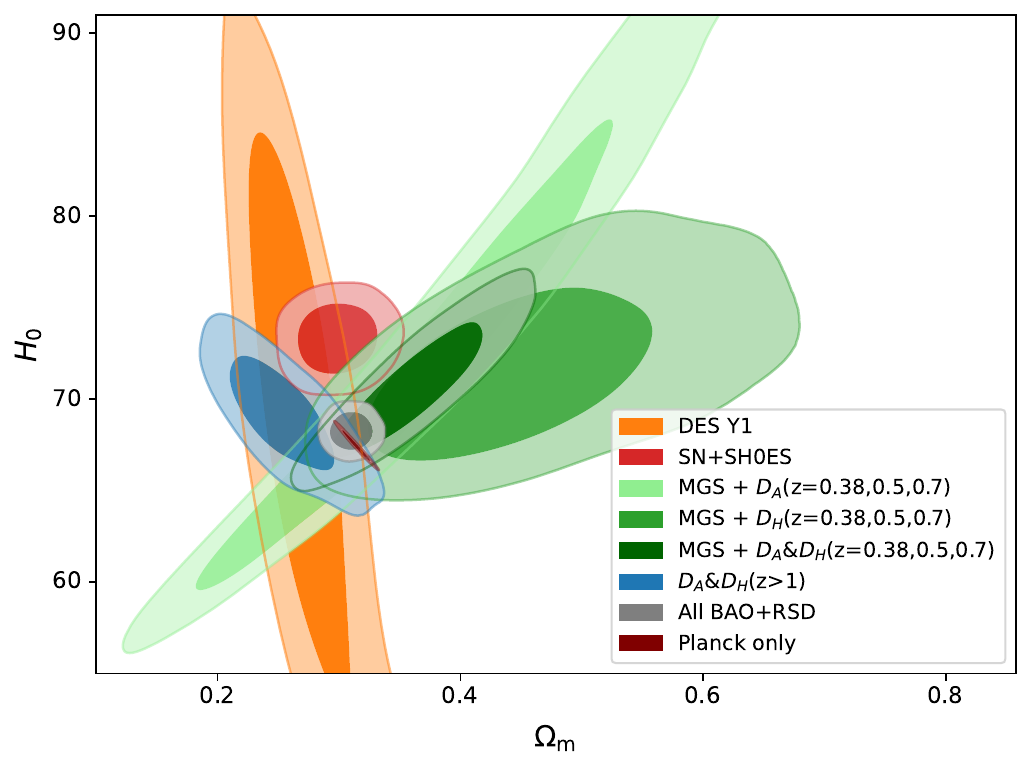}
 	\caption{Constraints of the $\Lambda$CDM in the $H_0 - \Omega_{\rm m}$ plane. The low- and high-$z$ BAO data do not point to the same region of the parameter space. The role of each individual data point on these contour can be understood by comparing to Fig.~\ref{fig:BAOindividuals}.}
 	\label{fig:BAO}
 	\vspace{-0.25cm}
 \end{figure}

The marginal constraints in the $H_0 - \Omega_{\rm m}$ plane for individual probes, obtained from an MCMC analysis, are gathered in Fig.~\ref{fig:BAO}. The Hubble tension is apparent in the fact that {\em Planck} only contours do not overlap with the ones from SN+SH0ES. However contours from {\em Planck} only intersect with those obtained from all BAO data only.

It is illuminating to explore in more details the BAO data by examining the constraints obtained from the separate high-$z$ and low-$z$ BAO datasets independently. It is clear on Fig.~\ref{fig:BAO} that this leads to rather different constraints in the $H_0 - \Omega_{\rm m}$ plane, a fact already highlighted in Fig.~5 of Ref.~\cite{eBOSS:2020yzd}. The low-$z$ BAO only data (without RSD) lead to a rather higher value of the matter content ($\Omega_{\rm m} = 0.388\pm0.050$), which is in slight tension with the DES Y1 constraints $\Omega_{\rm m}=0.248^{+0.030}_{-0.017}$  as well as with SN+SH0ES constraints $\Omega_{\rm m}=0.299\pm0.022$. Even though DES Y3~\cite{DES:2021wwk} now prefers larger values, this motivates that we either consider all BAO data or just the high-$z$ subset to assess our model.\\

\begin{figure}[htb]
 	\centering
 	\includegraphics[width=\columnwidth]{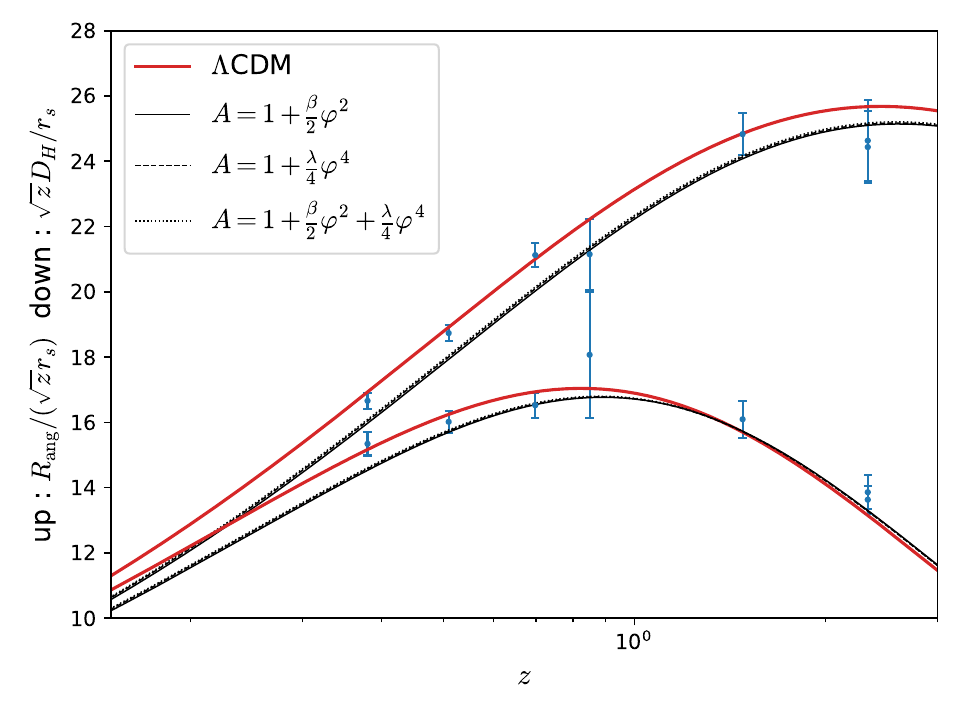}
 	\caption{Comparison of the BAO data and the best fits prediction for $D_H$ and $R_{\rm ang}$. These two quantities are related by Eq.~(\ref{eq:relDADH}) that tells us that $D_H$ is related to the slope of $R_{\rm ang}$ if the universe is described by a FL spacetime. It shows that the 3 data points with redshift  $z_{\rm eff}=0.38,0.5,0.7$ would prefer a mildly larger slope than the $\Lambda$CDM best fit. Note also that since our models generically predict  a lower $\Omega_{\rm m}$ they are in tension with the low-$z$ BAO data. Note also that even though the shapes of the transition in the 3 models differ slightly, they predict nearly identical post-recombination histories so that the curves for the 3 models cannot be distinguished on this plot. }
 	\label{fig:DADH}
 	\vspace{-0.25cm}
      \end{figure}

In addition, BAO observations provide an independent measurement of both $r_{\rm D} H(z)$ and $r_{\rm D}/R_{\rm ang}(z)$, with $r_{\rm D}$ the sound horizon at baryon drag ($z_{\rm D}\simeq 1060$) which is slightly larger than $r_s$, through their radial and angular measurements respectively. Indeed, in a spatially Euclidean FL spacetime,
\begin{equation}\label{eq:relDADH}
R_{\rm ang}(z) = \int_0^z D_H \dd z
\end{equation}
so that there should be a consistency between the $R_{\rm ang}$ and $D_H\equiv (c/H_0) E^{-1/2}(z)$ measurements, the latter being related to the slope of the former. This relation holds whatever the dynamics of the universe, i.e. independently of the matter content as long as it can be described by a FL spacetime. Fig.~\ref{fig:DADH}  shows that  the $R_{\rm ang}$ measurements  for the data points of redshifts $z_{\rm eff}=0.38,0.5,0.7$ tend to prefer a mildly larger slope than the $\Lambda$CDM best fit whereas the average slope set by the $D_H$ points in this redshift range is slightly smaller than the slope from this same best fit.

\begin{figure}[tb]
 	\centering
 	\includegraphics[width=\columnwidth]{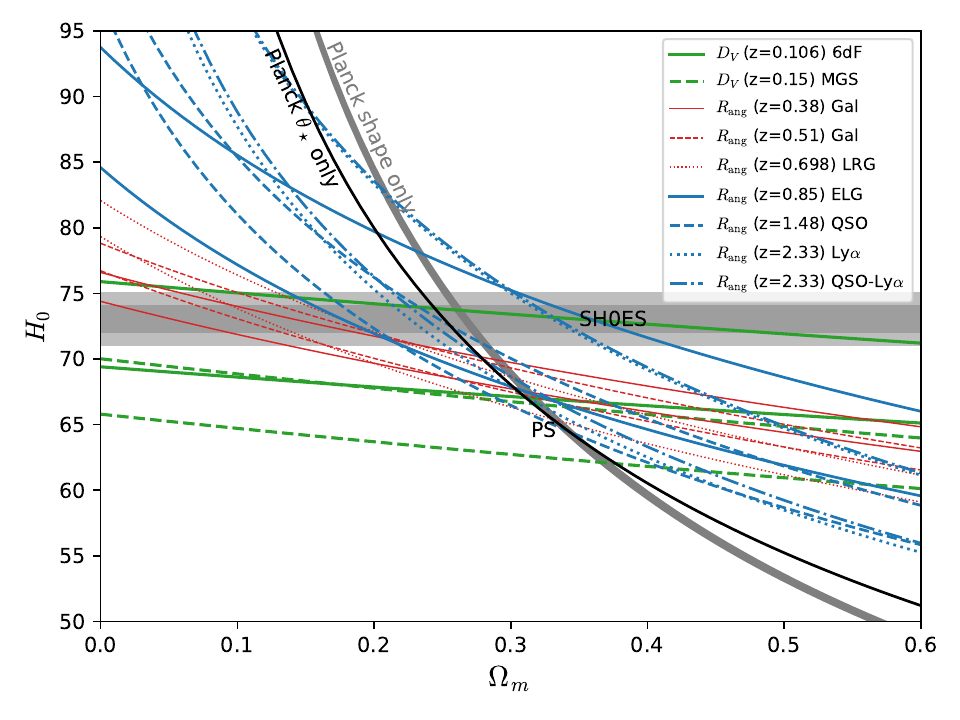}\\
        \includegraphics[width=\columnwidth]{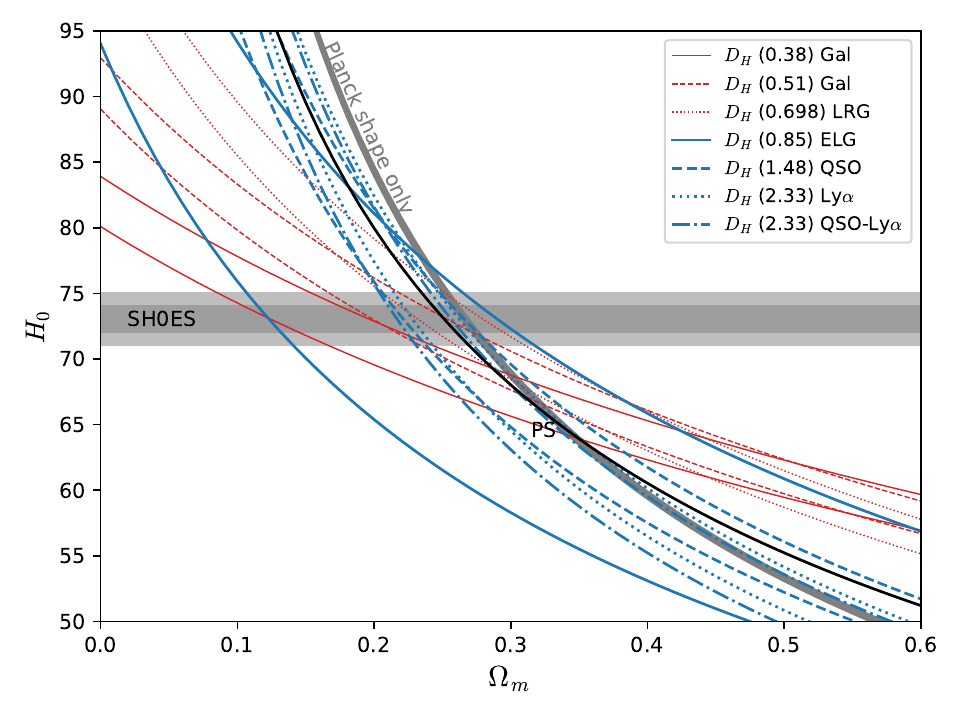}
 	\caption{Constraints from individual BAO data points. Assuming a fixed sound horizon from the $\Lambda$CDM {\em Planck} only best fit, one can determine the region ${\rm PS}$ allowed by the shape of the CMB angular spectrum: (1) its global shape (relative heights and positions of the acoustic peaks) sets the approximate constraint $\Omega_{\rm m} h^2 = 0.1442\pm 0014$ (gray line) while (2) the absolute positions of the acoustic peaks determine the angular size of the sound horizon $\theta_*$ (black line). Their intersection roughly determines the constraint ${\rm PS}$ set by {\em Planck} only. In order to be compatible with SH0ES, ${\rm PS}$ shall move to lower $\Omega_{\rm m}$. For each BAO points we depict the $1\sigma$ strip allowed by $R_{\rm ang}$ [{\it Top}] and $D_H$ [{\it Bottom}]. It is clear that if $H_0$ is determined by SH0ES, the $R_{\rm ang}$ data points at redshifts $z=0.38,0.5,0.7$ constrain $\Omega_{\rm m} \simeq 0.1$. The BAO data alone constraints drags towards a larger $\Omega_{\rm m}$.}
 	\label{fig:BAOindividuals}
 	\vspace{-0.25cm}
 \end{figure}

To better understand these slight tensions, it is instructive to detail the constraints in the $H_0-\Omega_{\rm m}$ plane for each individual BAO measurements. Fig. \ref{fig:BAOindividuals} summarizes these individual constraints assuming  $r_{\rm D}=147.07\,{\rm Mpc}$ from the Planck $\Lambda$CDM model best fit (in the $\Lambda$CDM model); see appendix~\ref{app:DADH} and Eq.~(\ref{e.rD}) for the expression of $r_{\rm D}$. This differs from Fig.~\ref{fig:BAO}  in which the constraints take into account the $(H_0,\Omega_{\rm m}$)-dependence of the sound horizon, approximately as $r_s \propto 1/(H_0\Omega_{\rm m}^{1/2})$ and similarly for $r_{\rm D}$.  Any change in the constant $r_{\rm D}$ chosen would shift the constraints of Fig. \ref{fig:BAOindividuals} vertically.

Assuming a fixed $r_{\rm D}$, it is clear that the $R_{\rm ang}$ measurements for the 3 data points with redshifts $z=0.38,0.5,0.7$ are in full agreement with the MGS constraint on $D_V\equiv (z R_{\rm ang}^2 D_H)^{1/3}$ at $z=0.15$ {\em only }for large values of $\Omega_{\rm m}$.  The $D_H$ data at these same 3 redshifts do also agree well only for large values of $\Omega_{\rm m}$. However such large  $\Omega_{\rm m}$ are in tension with the SN data. It follows that any early modified cosmological model that would induce a lower $r_{\rm D}$, will shift all curves upwards hence bringing them in a better agreement with the $H_0$ constraint by SH0ES. But their agreement would still be for the same large values of $\Omega_{\rm m}$, as seen in Fig. \ref{fig:BAO}. Hence, a tension with SN data would persist. Conversely, BAO data from redshifts $z=0.85,1.48,2.33$, be it $R_{\rm ang}$ or $D_H$, are in agreement for values of $\Omega_{\rm m}$ which are roughly around $0.3$ and which are in better agreement with SN data.

This issue can be grasped differently by considering the intersection of the BAO constraints with the SH0ES constraint on $H_0$ while still assuming the sound horizon at baryon drag $r_{\rm D}=147.07\,{\rm Mpc}$. The $R_{\rm ang}$ data points at redshifts $z=0.38,0.5,0.7$ would constrain $\Omega_{\rm m} \simeq 0.1$ whereas the rest of the BAO data would indicate $0.2 \lesssim \Omega_{\rm m} \lesssim 0.3$, exception made of the 6dF and MGS data point at $z=0.15$, which cannot accommodate the SH0ES result in any manner with such a sound horizon. 

Therefore, either a solution arises from a modification of the sound horizon at recombination while preserving the quality of the fit to the CMB peaks, but this can only be partially successful since the sound horizon rescaling needed for low-$z$ BAO is necessarily different~\cite{Jedamzik:2020zmd}, or without any modification to the sound horizon the BAO data for $R_{\rm ang}$ at low-$z$ would hold the constraints of any alternative model to intermediate values of  $H_0$, between the Planck and the SH0ES values, and thus cannot fully resolve the Hubble tension. This is a limit imposed by the current structure of the data that no model can beat. For all these reasons, we have decided to also consider a subset of BAO data with all BAO and RSD data, except those three values of $R_{\rm ang}$ at $z_{\rm eff}=0.38,0.5,0.7$ and the data points from the MGS at $z=0.15$, but keeping all $D_H$ and RSD with their correlated errors at these redshifts. This BAO subset is noted hereafter BAO(-4pts)+RSD.

To finish, let us come back on the informations that can be extracted from the global shape of the CMB angular power spectrum, that is the relative heights of the acoustic peaks and their angular positions. First, note that the relative heights of the peaks and their relative positions, regardless of the sound horizon angle $\theta_\star$, constrain the relative ratios of DM to photons and to baryons around recombination. In the $\Lambda$CDM, and therefore knowing the dilution of all species, it provides information on $\Omega_{\rm D0} h^2$ since today the photon energy density is well known and the redshift of recombination is also well constrained. To get a crude estimate of the constraints it sets in the $H_0 - \Omega_{\rm m}$ plane, we can combine the marginal from {\em Planck} on $\Omega_{\rm D0} h^2$ to the one on $\Omega_{\rm b0} h^2$ to get the rough constraint $\Omega_{\rm m} h^2 \simeq 0.1442\pm 0.0014$ from the CMB global shape alone. We include the corresponding region as a gray curve in Fig.~\ref{fig:BAO}. In addition we can also add the constraint arising from the angular size of the sound horizon $\theta_\star$ which is very well measured from the absolute positions of the CMB peaks (black curve). Their domain of intersection ${\rm PS}$ in the $H_0 - \Omega_{\rm m}$ plane roughly determines the constraint set by {\em Planck} data alone: it implies a value for $H_0$ too low to agree with SH0ES. Any solution that solves the Hubble tension while preserving the sound horizon at recombination, would also preserve the sound horizon at baryon drag ($r_{\rm D} =147.07\,{\rm Mpc}$ for the $\Lambda$CDM best fit). Therefore, it necessarily requires to find a way to shift the constraints from the CMB global shape, i.e. the gray curve, to the left so that the new ${\rm PS}$ domain lies within the region compatible with SH0ES. This is actually what our models do by triggering the disappearance of DM around and after recombination. This implies that such class of solutions must generically have a lower $\Omega_{\rm m}$ than the $\Lambda$CDM best fit, and consequently a larger cosmological constant. Generically, in our models, the universe expands slower after recombination until $\Lambda$ starts to dominate (and it dominates earlier), at which point it expands faster. This leads to the same comoving distance to the LSS but with a higher $H_0$ and a younger universe.
 
\subsection{MCMC analysis}

\begin{figure}[tb]
 	\centering
	 \includegraphics[width=\columnwidth]{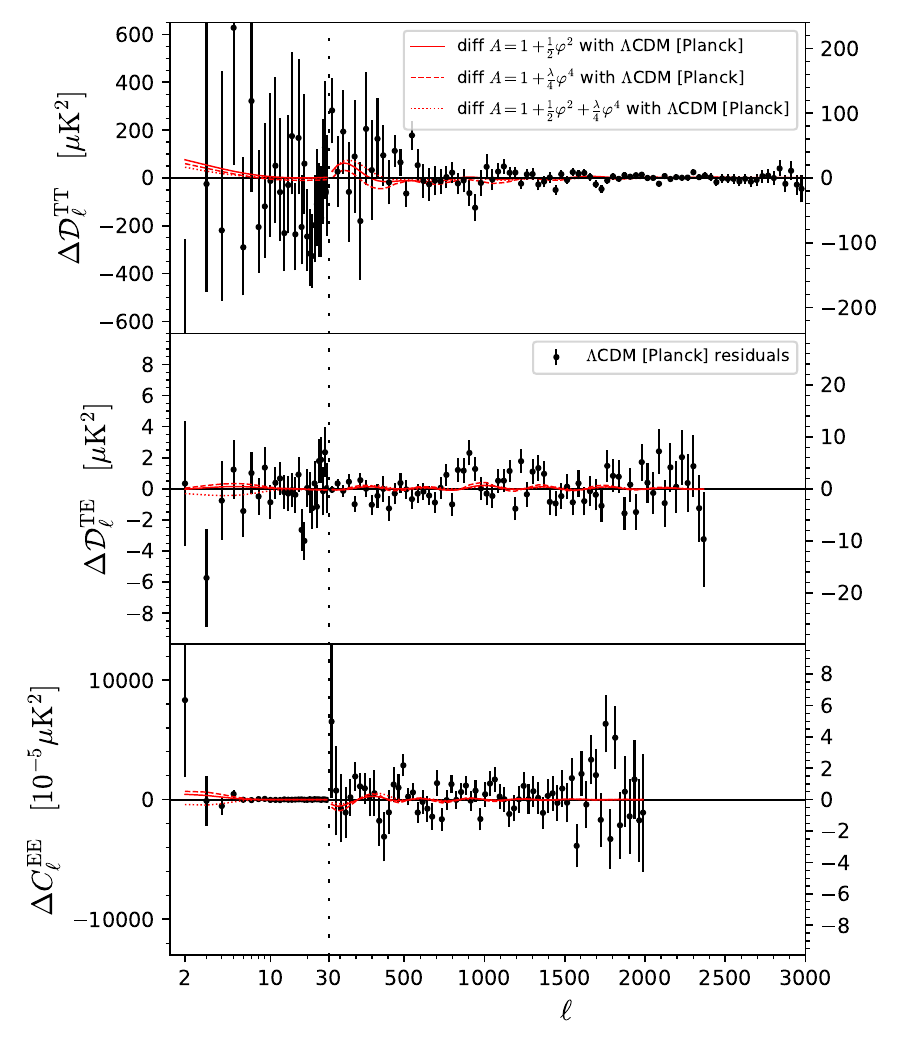}
		\vspace{-0.75cm}
 	\caption{{\it Black:} residuals of TT, TE and EE spectra of the $\Lambda$CDM best fit (from Planck data only)~\cite{Planck:2018vyg} with Planck data. {\it Red:} Differences between our $\Lambda\beta$CDM best fit spectra (obtained with the base dataset+BAO($z>1$)+$H_0$ and the former spectra.}
 	\label{fig:residuals}
 	\vspace{-0.25cm}
\end{figure}

\begin{figure}[tb]
 	\centering
 	\includegraphics[width=\columnwidth]{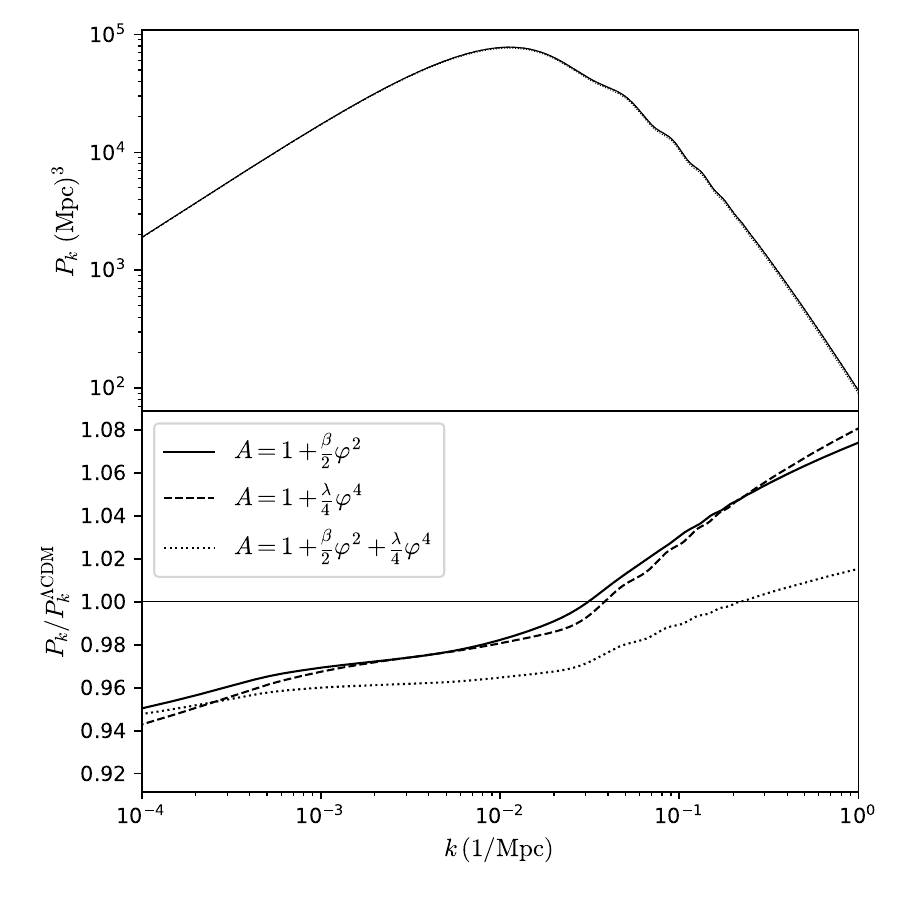}
 	\caption{[{\it Top}] The power spectra of the best fits of the 3 models are indistinguishable by eye. [{\it Bottom}] Relative variation of the final matter power spectrum with respect to the $\Lambda$CDM best fit. It is mostly boosted on smaller scales beyond the peak by less that 10\%. This implies that $S_8$ is almost unaffected.}
 	\label{fig:Pkratio}
 	\vspace{-0.25cm}
 \end{figure}

We can now present the results of the full MCMC comparisons for the $\Lambda\beta$CDM, $\Lambda\lambda$CDM and $\Lambda(\beta,\lambda)$CDM models. All cosmological parameters are varied in a spatially Euclidean cosmology, along with the new parameters $\beta$, $\lambda$ and the attractor initial value $\varphi_i$ of the scalar field. We use large flat prior for the parameters of the $\Lambda$CDM model and flat priors for the scalar field with $A'(\varphi_i)/\varphi_i \in [0.2,1]$ (hence translating into a prior on $\beta$ for the $\Lambda\beta$CDM model), and $A(\varphi_i) \in [1, 1.1]$. In addition, in the $\Lambda(\beta,\lambda)$CDM model we use the flat prior $\lambda/\beta^2 \in [-1,0.5]$. The neutrino masses are taken into account by assuming that one flavour of massive neutrinos with $m_\nu = 0.06\,{\rm eV}$, and setting the number of massless relativistic species so as to ensure $N_{\rm eff} = 3.044$~\cite{Froustey:2020mcq}.

It can  first be checked from Fig.~\ref{fig:residuals} that the best fits for all models fit well the CMB patterns as expected from earlier discussions. Their residuals with the CMB data are as good as those of the $\Lambda$CDM best fit. For these best fits, the preferred parameters correspond to sound horizons which are extremely similar to the $\Lambda$CDM best fit within the same set of data, ensuring that the $\chi^2$ for CMB data is not degraded significantly even though our priors allow our model to have a modified sound horizon. Then, Fig.~\ref{fig:Pkratio} provides some insight on the matter power spectrum. Since the modes growth is almost unaffected, it is very similar to the $\Lambda$CDM power spectrum below its peak while it has slighlty more power on small scales beyond the peak. Since $\Omega_{\rm m}$ is decreased by the coupling with the scalar field, $S_8$ increases only marginally,  which is a positive feature of this class of models. 

 \begin{figure*}[!htb]
 	\includegraphics[width=0.95\textwidth]{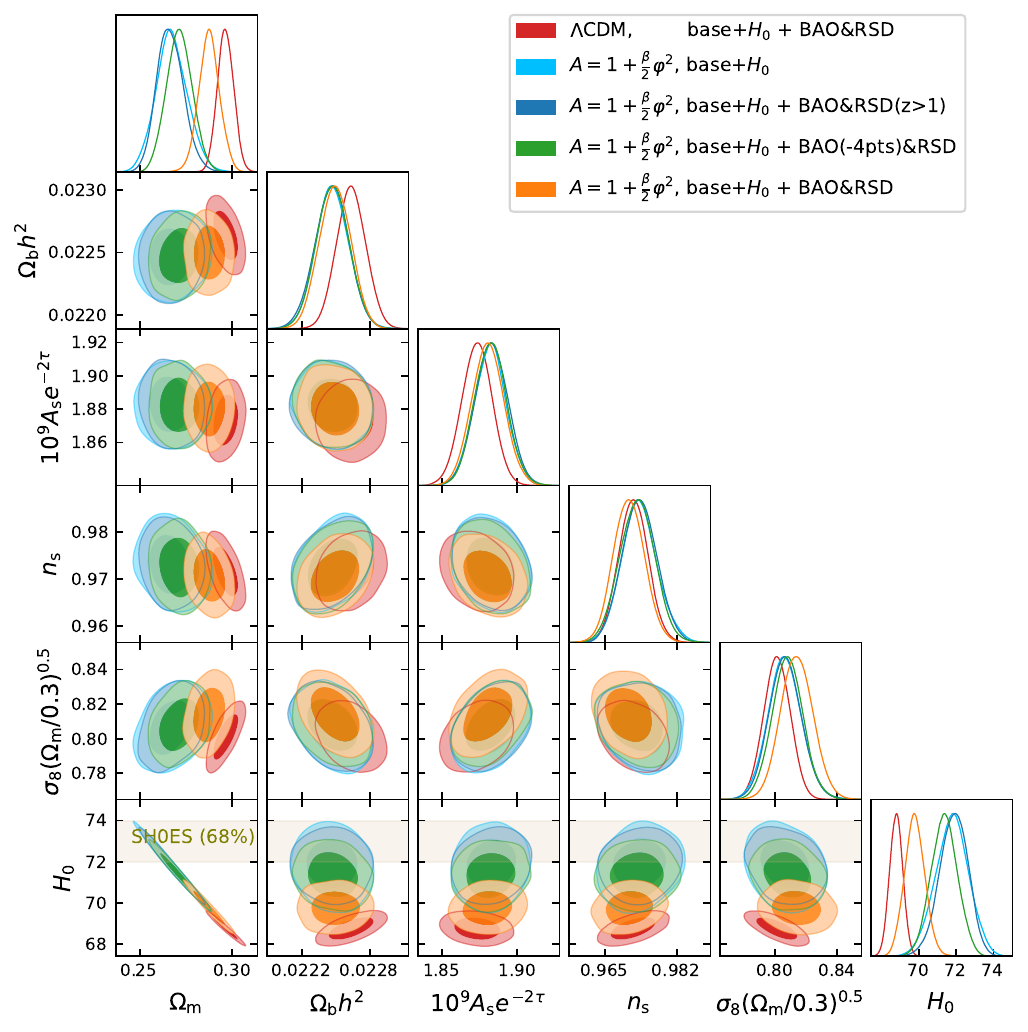}
 	\caption{Marginal distributions of the $\Lambda\beta$CDM model. The $\Omega_{\rm m}-H_0$ degeneracy is clearly seen on the lower left panel. The model moves along this line toward a lower $\Omega_{\rm m}$ to get a higher $H_0$. All other cosmological parameters are almost unaffected and remain close to their $\Lambda$CDM value but $\Lambda$ that will adjust since we have assumed a spatially Euclidean cosmology ($K=0$).}
 	\label{fig:triangleharmonicfull}
\end{figure*}

\begin{figure*}[htb]
        \includegraphics[width=0.9\columnwidth]{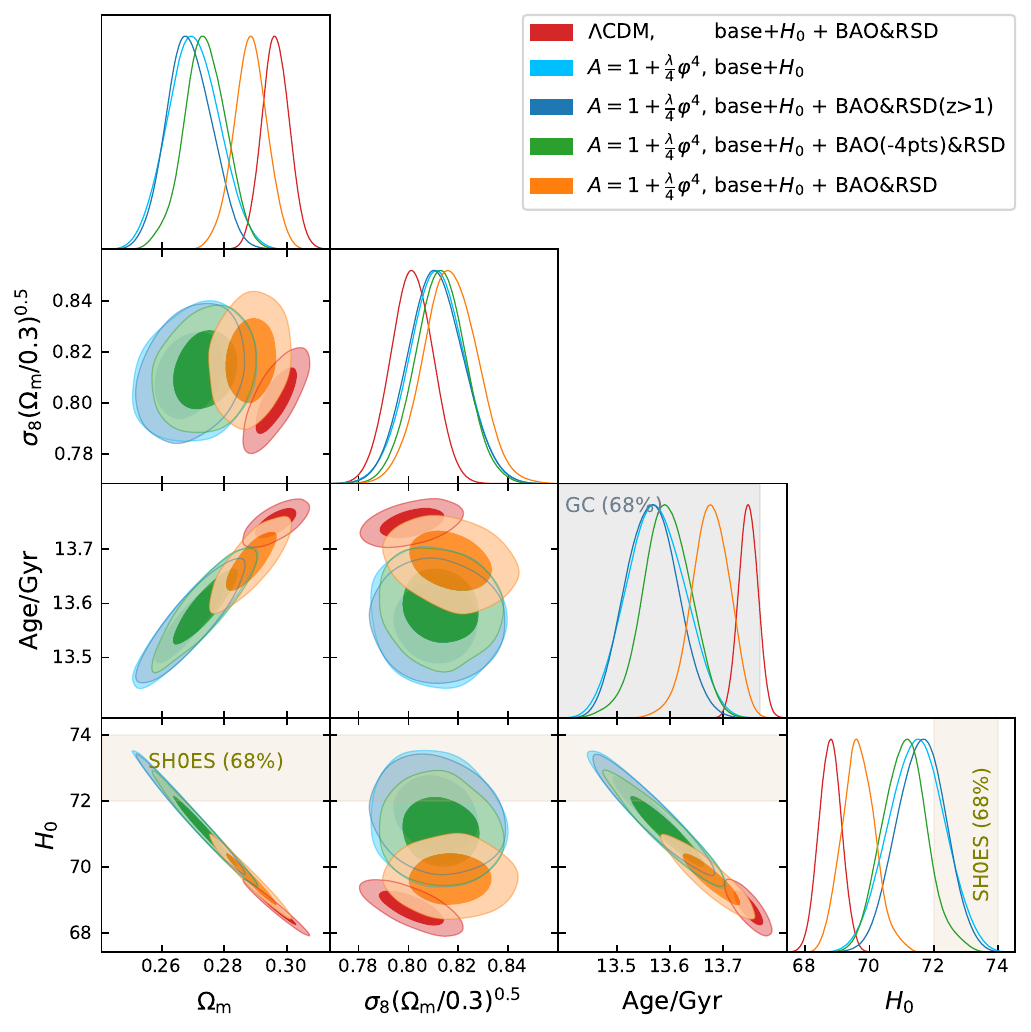}\includegraphics[width=0.9\columnwidth]{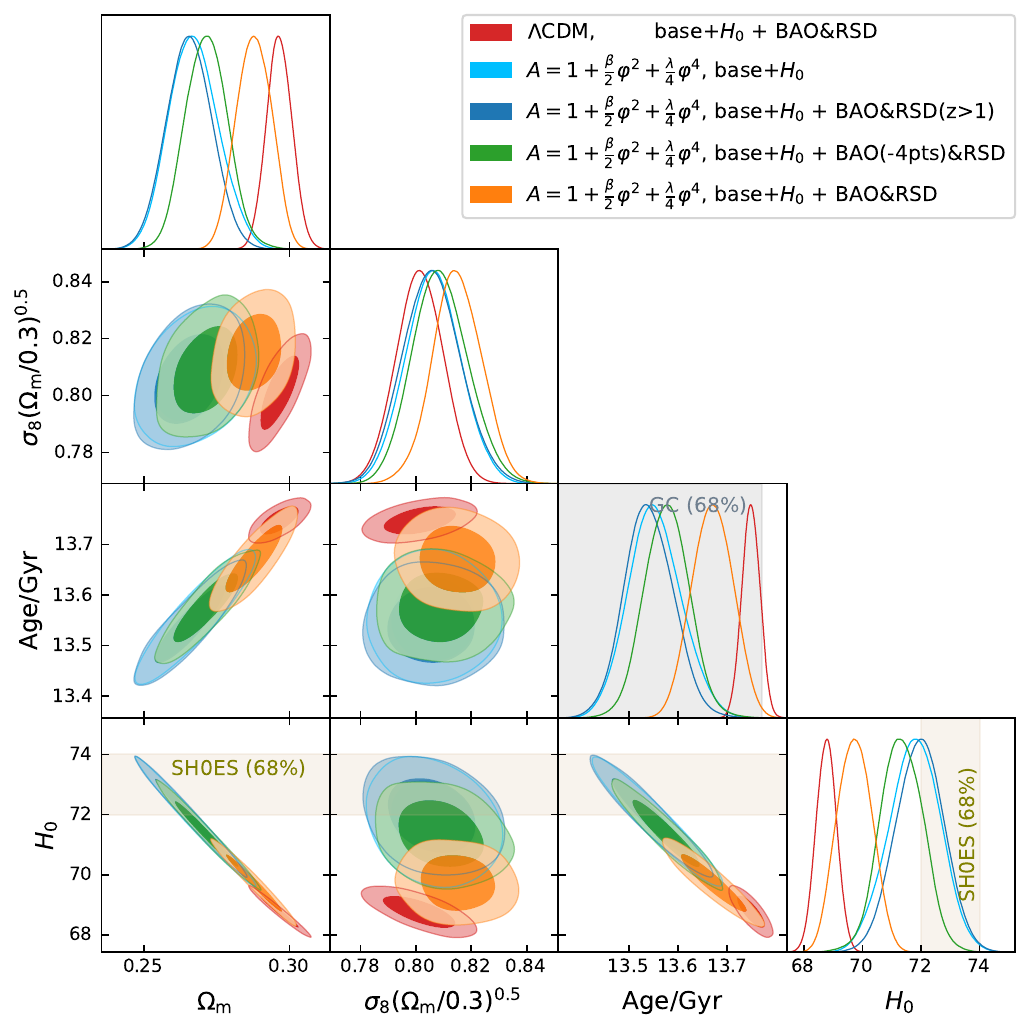}
 	\caption{Marginal distributions of the $\Lambda\lambda$CDM and $\Lambda(\beta,\lambda)$CDM models. All models share the same properties to lead to a lower $\Omega_{\rm m}$ to allow for a higher $H_0$ generically concluding for a younger universe compared to the $\Lambda$CDM and $S_8$ remains almost unchanged.}
 	\label{fig:tri23}
 	\vspace{-0.25cm}
  \end{figure*}

The $\Lambda\beta$CDM being the simplest model, we start by discussing the marginal distributions of all its relevant cosmological parameters presented in Fig.~\ref{fig:triangleharmonicfull}, with the different combinations of BAO data detailed above. For the two other models  $\Lambda\lambda$CDM and $\Lambda(\beta,\lambda)$CDM, we only show the marginal distributions on a reduced set of parameters  in  Fig.~\ref{fig:tri23}.  All models share the same properties of a lower $\Omega_{\rm m}$ as expected, a younger universe and a lower tension with SH0ES on $H_0$. The constraints on the cosmological parameters are gathered in Table~\ref{tab:1} and we summarized the best fit parameters of the four models in Table~\ref{tab:2}. They lead to similar constraints, all sharing the same trends that we discuss in more details for the $\Lambda\beta$CDM.

  \begin{table*}[htp]
    \centering
    \resizebox{\textwidth}{!}{
\begin{tabular}{|ll|ccccc|ccc|}
\hline
Model & base+$H_0$+& $\Omega_{\rm m}$ & $\Omega_{\rm b0} h^2$ & $h$  & $S_8$   &  Age (Gyr) & {\rm GT}  & $Q_{\rm DMAP}$ &$\Delta$AIC \\
  \hline
$\Lambda$CDM & BAO &   $0.2965\pm0.0044$ & $0.02263\pm0.00013$   &  $0.6877\pm0.0035$ & $0.801\pm0.009$ & $13.75\pm0.02$  & 4.4$\sigma$& 4.8 & 0 \\
$\Lambda$CDM & BAO($z>1$) &   $0.2912\pm0.0052$ & $0.02270\pm0.00014$   &  $0.6919\pm0.0042$ & $0.794\pm0.010$ & $13.73\pm0.02$  & 4.1$\sigma$& 4.4 & 0 \\
$\Lambda\beta$CDM & BAO &   $0.2875\pm0.0056$ & $0.02249\pm0.00014$   &  $0.6977\pm0.0054$ & $0.814\pm0.010$ & $13.67\pm0.04$  & 3.8$\sigma$& 3.6 & -2.6 \\
$\Lambda\beta$CDM & BAO($z>1$) &   $0.2666\pm0.0073$ & $0.02246\pm0.00015$   &  $0.7187\pm0.0076$ & $0.807\pm0.010$ & $13.55\pm0.05$  & 1.8$\sigma$& 2.0 & -14.5 \\
$\Lambda\lambda$CDM & BAO &   $0.2884\pm0.0054$ & $0.02253\pm0.00014$   &  $0.6966\pm0.0052$ & $0.817\pm0.011$ & $13.68\pm0.04$  & 3.8$\sigma$& 3.9 & -2.8 \\
$\Lambda\lambda$CDM & BAO($z>1$) &   $0.2689\pm0.0072$ & $0.02253\pm0.00014$   &  $0.7160\pm0.0075$ & $0.811\pm0.011$ & $13.56\pm0.047$  & 2.2$\sigma$& 2.0 & -15.4 \\
$\Lambda(\beta,\lambda)$CDM & BAO &   $0.2878\pm0.0059$ & $0.02247\pm0.00015$   &  $0.6974\pm0.0058$ & $0.815\pm0.009$ & $13.67\pm0.04$  & 3.8$\sigma$& 3.5 & -0.1 \\
$\Lambda(\beta,\lambda)$CDM & BAO($z>1$) &   $0.2656\pm0.0077$ & $0.02246\pm0.00015$   &  $0.7196\pm0.0081$ & $0.805\pm0.010$ & $13.54\pm0.04$  & 1.7$\sigma$& 1.8 & -13.8 \\
  \hline 
\end{tabular}}
\caption{Comparison of the posterior marginal distributions and success criteria of the $\Lambda$CDM, $\Lambda\beta$CDM, $\Lambda\lambda$CDM and $\Lambda\beta\lambda$CDM models. GT is the Gaussian tension on $H_0$ for the model between the dataset mentioned (but without $H_0$),  and the $H_0$ constraint from SH0ES only. Since spatial sections are Euclidean and radiation energy density is negligible today, the energy fraction of the cosmological constant is constrained by $\Omega_\Lambda + \Omega_{\rm m} = 1$. \label{tab:1}}
\end{table*}

With the base+BAO+$H_0$ dataset, all models alleviate the tension with SH0ES since for a $\Lambda$CDM the average Hubble constant is $\bar h=0.688$ whereas in the  $\Lambda \beta$CDM it is $\bar h=0.698$. This means that the 4.4$\sigma$ tension is reduced to 3.8$\sigma$. We consider a measure of the tension within a given model defined as (see Refs.~\cite{Schoneberg:2021qvd,Khalife:2023qbu} for the methodology)
\begin{equation}
Q_{\rm DMAP}  = \sqrt{\chi^2_{{\rm dataset+}H_0} - \chi^2_{{\rm dataset}}}
\end{equation}
where $\chi^2$ stands for the minimum of $-2\ln({\cal L})$ (obtained with BOBYQA~\cite{2018arXiv180400154C,2018arXiv181211343C}). $Q_{\rm DMAP}$ measures the degradation of the best fit within a model when the $H_0$ constraint from SH0ES is included. Also in order to compare the merit of a given model with respect to the $\Lambda$CDM model, we consider
\begin{equation}
\Delta{\rm AIC} = \chi^2_{{\rm model, dataset+}H_0} - \chi^2_{\Lambda{\rm CDM, dataset+}H_0} + 2 N
\end{equation}
where $N$ is the number of additional parameters of the model (2 for $\Lambda\beta$CDM and $\Lambda\lambda$CDM, but 3 for $\Lambda(\beta,\lambda)$CDM).

For the $\Lambda \beta$CDM model, and with dataset=base+BAO, we find $Q_{\rm DMAP} = 3.6$  and $\Delta {\rm AIC} =- 2.6$, which indicates only a marginal improvement. However, as we have discussed in the previous section, the low-$z$  BAO data favor a higher matter fraction today.
As a consequence, when excluding 4 points of the low-$z$ BAO or the whole $(z<1)$-BAO datasets [that is considering respectively dataset=base+BAO(-4pts) and dataset=base+BAO($z>1$)], the model can reach a higher $H_0$ since lower $\Omega_{\rm m}$ values are allowed. With the BAO($z>1$), it leads to $h=0.7187\pm0.0076$, in good agreement with SH0ES, while other criteria also improve substantially ($Q_{\rm DMAP} =2.0$ and $\Delta {\rm AIC} =-14.5$). The constraints with BAO(-4pts) are very similar, as can be checked on Fig.~\ref{fig:triangleharmonicfull}, showing that the four low-$z$ BAO data points have the strongest statistical weight toward a lower $H_0$. Note also that within the low-$z$ BAO data, only the ones measuring $R_{\rm ang}$ show this trend as the $D_H$ and RSD measurements at the low redshifts are kept in the BAO(-4pts) dataset.

In the constraints with the BAO($z>1$) dataset, the universe is younger with $t_{\rm U} = 13.55\pm0.05\,{\rm Gy}$, an age consistent with the one deduced from globular clusters (GC) \cite{Valcin:2021jcg,Bernal:2021yli}. Note also that $S_8 = 0.807\pm 0.010$, hence the tension with DES results increases only mildly.  To finish, note that the small tension on $\Omega_{\rm b}h^2$ mentioned in Refs.~\cite {Pitrou:2020etk,Pitrou:2021vqr} survives since these models are precisely built to avoid any alteration of BBN physics. 

Our analysis also emphasized on a concrete example the problem of reconciling CMB data with low-$z$ BAO data, as pointed out in Ref.~\cite{Jedamzik:2020zmd}. This confirms the insight~\cite{Vagnozzi:2023nrq} that ``new physics is not sufficient to solve the $H_0$ problem'', see also Refs.~\cite{Addison:2017fdm,Keeley:2022ojz,Efstathiou:2021ocp,Cai:2021weh,Aylor:2018drw} for similar arguments.

\begin{table*}[htp]

\begin{center}
\begin{tabular}{|ll|cccccc|ccc|}
\hline
Model & base+$H_0$+& $\Omega_{\rm m}$ & $\Omega_{\rm b0} h^2$ & $h$  & $\log\left(10^{10}A_s\right)$ & $n_s$ & $\tau_{\rm reio}$ & $\beta$ & $\lambda$ & $\varphi_i$ \\
\hline
$\Lambda$CDM & BAO &  0.3011 & 0.02259  & 0.6841 & 3.0524 & 0.9674 & 0.0598 & 0 & 0 & 0 \\
$\Lambda$CDM & BAO($z>1$) & 0.2957  & 0.02261  & 0.6886 & 3.0652 & 0.9735 & 0.0666 & 0 & 0  & 0 \\
$\Lambda\beta$CDM & BAO &  0.2836  & 0.02246  & 0.7012 & 3.0454 & 0.9720 & 0.0542 & 0.3196 & 0 & 0.4345 \\
$\Lambda\beta$CDM & BAO($z>1$) & 0.2619  & 0.02250  & 0.7240 & 3.0569 & 0.9741 & 0.0601 & 0.2509 & 0 & 0.7159 \\
$\Lambda\lambda$CDM & BAO & 0.2863  & 0.02242  & 0.6987 & 3.0544 & 0.9708 & 0.0591 & 0 & 0.8436 & 0.6551 \\   
$\Lambda\lambda$CDM & BAO($z>1$) & 0.2659  & 0.02253  & 0.7188& 3.0625 & 0.9769 & 0.0647 & 0 & 0.1666  & 1.0959 \\   
$\Lambda(\beta,\lambda)$CDM & BAO & 0.2870  & 0.02249  & 0.6979 & 3.0409 & 0.9713 & 0.0518 & 0.2862 & -0.0653 & 0.4242 \\   
$\Lambda(\beta,\lambda)$CDM & BAO($z>1$) & 0.2681  & 0.02248  & 0.7167 & 3.0306 & 0.9713  &0.0477  & 0.2551  & -0.0426 & 0.6489 \\   
  \hline 
\end{tabular}
\caption{Best fits of  $\Lambda$CDM, $\Lambda\beta$CDM, $\Lambda\lambda$CDM and $\Lambda(\beta,\lambda)$CDM models. All the figures of this article have been performed with these values obtained using the BAO($z>1$) datasets.
\label{tab:2}}
\end{center}
\end{table*}%

\section{Conclusions}\label{secconlc}

The theory presented in this article and the cosmological models that can be constructed on its basis are simple and minimal extensions of the $\Lambda$CDM in the sense that the physics of the visible SM sector remains fully unchanged. We take advantage of the fact that the physics of the dark sector remains badly constrained. We assume that DM enjoys a long-range scalar interaction besides the standard GR interaction, i.e. a dark scalar force. The main direct consequences of these assumptions are:
\begin{enumerate}
\item No constants that can be measured in the laboratory or in the Solar system vary. This implies that it escapes {\em by construction} all existing local constraints on the deviation from GR in the Solar system. There is no testable violation of the weak equivalence principle~\cite{MICROSCOPE:2022doy} or observable variation of a non-gravitational constant~\cite{Uzan:2002vq,Uzan:2010pm} since they concern the SM fields.
\item The gravitational constant remains constant and equal to the Cavendish value of the Newton constant since, again, SM fields are not subjected to the scalar interaction.
\item The energy density of the new scalar degree of freedom is subdominant in the energy budget during the whole cosmic history (see Fig.~\ref{fig:2}). Hence this is not a model of dark energy; $\varphi$ does not affect directly the expansion history of the universe and cannot explain the acceleration of the cosmic expansion. 
\item From the two previous points, it is obvious that BBN predictions remain fully unaffected {\em by construction}.
\item It also implies that the interpretation of the Hubble diagram remains unaffected.
\item The gravitational effect of DM on the SM field is unchanged since the latter have no charge under the new interaction.
\item But the gravitational clustering of DM is affected. As we have seen, in the cosmological setting, the non-minimal coupling can be interpreted either as a variation of the DM gravitational constant or as the fact that part of the DM energy density as been transfered to $\rho_\varphi$. As a consequence DM gets an effective non-vanishing time-varying equation of state (see Fig.~\ref{fig:wphi}).
\end{enumerate}

 A fully consistent theory has been proposed in Eqs.~(\ref{e.TH1}-\ref{e.TH5}) assuming a massless scalar field and a coupling~(\ref{e.tildeg}) to DM. This allows us to treat in a consistent way the background and perturbation evolutions and to ensure that the non-cosmological constraints are also taken into account. Actually, among the side results, we have illustrated the effect of not taking the perturbations consistently into account by considering a phenomenological model with same background cosmology but without the effect of the non-minimal coupling at the perturbation level. This led to errors of order of several percents on the CMB predictions (see Fig.~\ref{fig:Clwrong}) which is much higher that what the data can allow. We have also discussed the reconstruction problem of models with time varying dark energy and dark matter equations of states, or equivalently DM couple to DE models.  We have also stressed the importance to take into account non-cosmological data. Indeed the comparison to data can state whether  a model is excluded at a certain confidence level but comparing best fits cannot rely only on data. While  one can indeed compare their merit to fit a given dataset, this is not sufficient, and one shall also attribute a prior on the theoretical construction and consistency of the theory that reflects both our prior beliefs~\cite{Trotta:2008qt,Trotta:2017wnx} and the ability of the theory to make predictions or not in any physical situation.  In particular, models assuming e.g. ad hoc phenomenological evolution of background quantities or constants, i.e. solutions to an unknown theory, shall have a much lower credence than a theory to which one finds a consistent solution in any physical situation.
 
The models studied in this article offer an effective way to make the gravitational constant in the dark sector vary with cosmic time or equivalently to have a suppression of the DM energy density right after the onset of the matter era. Interestingly, our models have the key generic feature that they let the sound horizon and distance to the last scattering surface unchanged compared to the standard $\Lambda$CDM but with a higher $H_0$ at the expense of a lower $\Omega_{\rm D0}$, which is still in agreement with low-$z$ data and in particular DES Y1. This is the key feature of these models that explains why they improve the Hubble tension (see Fig.~\ref{fig:BAOindividuals}). All our models predict a younger universe than the $\Lambda$CDM. Among the models compared here, the quadratic~(\ref{e.defmodel1}) and quartic~(\ref{e.defmodel2}) couplings offer a similar quality of fit, while adding a quartic contribution to the quadratic one~(\ref{e.defmodelE}) does not improve it. Typically with all data the Hubble tension drops from 4.4$\sigma$ to 3.8$\sigma$, while it is reduced from 4.1$\sigma$ to 1.8$\sigma$ when low-$z$ BAO are not used. Again, we recall that our models drive $\Omega_{\rm m}$ to lower values. This led us to discuss the role of the high- and low-$z$ BAO data, a key issue to hope to fully solve the Hubble tension, but indeed too far from the main scope of this work. While in agreement with DES-Y1, we have to emphasize that DES Y3 results \cite{DES:2021wwk} are shifted upwards with $\Omega_{\rm m}=0.339^{+0.032}_{-0.031}$, which can put that solution to the Hubble tension at risk. Also it may be challenged by the Pantheon+ \cite{Brout:2022vxf} sample which finds $\Omega_{\rm m}=0.334\pm 0.18$. As pointed out in \cite{Blanchard:2022xkk}, these large recent values of $\Omega_{\rm m}$ imply that most extensions of the $\Lambda$CDM model aiming at solving the Hubble tension lead to a tension on $\Omega_{\rm m} h^2$.

\begin{figure}[htb]
        \includegraphics[width=0.45\textwidth]{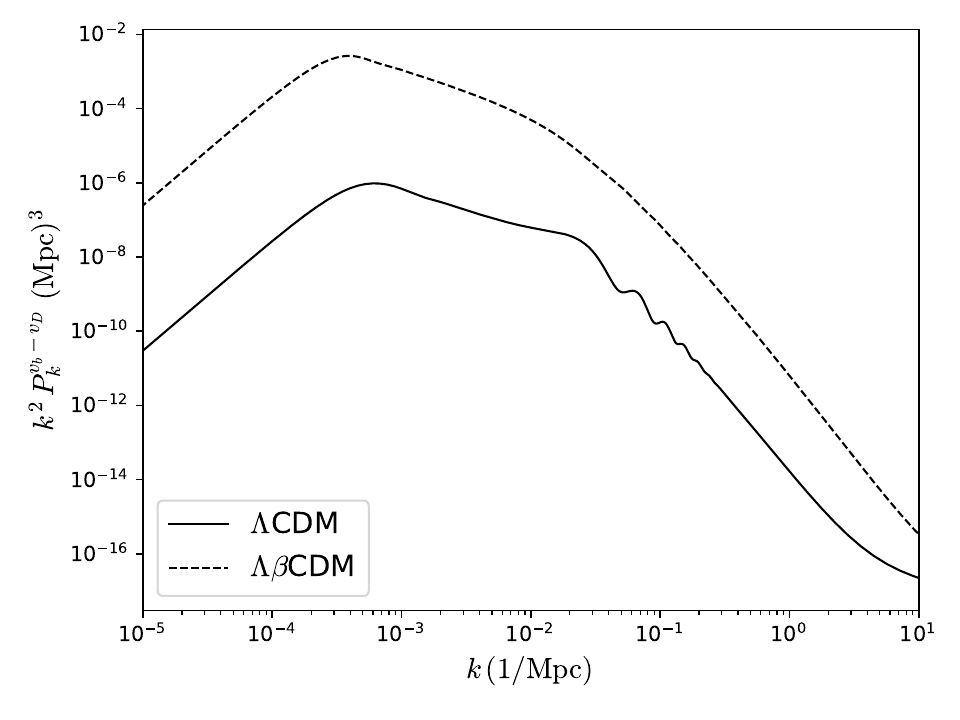}
 	\caption{Power spectra of the baryon-CDM velocity difference $k(v_{\rm b} - v_{\rm D})$ in the $\Lambda$CDM and $\Lambda\beta$CDM models.}
 	\label{figvbc}
 	\vspace{-0.25cm}
  \end{figure}
  
Our model is almost indistinguishable from the $\Lambda$CDM. It enjoys however the specific feature that DM and baryons do not feel the same gravity which will imprint their velocity. In Fig.~\ref{figvbc} we compared the power spectrum of $k(v_{\rm b}-v_{\rm D})$ in the  $\Lambda\beta$CDM model to the standard case. It shows two specific features: (1) it is boosted by 4
orders of magnitude and (2) the baryonic oscillations are washed out. It follows that this would modify the measurement of velocity induced acoustic oscillations~\cite{Tseliakhovich:2010bj,Schmidt:2016coo} that could be measured from 21cm observations~\cite{Munoz:2019rhi,Munoz:2019fkt}. This observable will be sensitive to any effect biasing the velocity of the two kinds of matter, and hence a possible new probe of the equivalence principle between the visible and dark sectors. Similarly, Refs.~\cite{Bonvin:2018ckp,Bonvin:2020cxp,Umeh:2020cag,Bonvin:2022tii} proposed to test the WEP in the dark sector by combining redshift-space distortions and relativistic effects in the galaxy number-count fluctuations, which could be achieved with the coming DESI and SKA observations. Both require a dedicated study.

The minimal models consider a single extra-function $A(\varphi)$ for a massless dark sector dilaton.  They can be extended in many ways by:
\begin{enumerate}
\item considering different coupling functions $A$. We have compared 3 of them here and shown that there signatures only differ slightly. The variation of $A$ from its initial to its lowest value controls the amplitude of the effect on DM, whereas the parameter $K_0$ defined in Eq. \eqref{eq:defK0b}, and which is related to the initial slope of the function, controls when the effect takes place;
\item introduce a non-vanishing potential $V$. We restricted to a massless $\varphi$. A potential can modify the cosmological dynamics and in particular make it a quintessence field if it dominates at late time; Since the mass of the quintessence field should be of the order of the late-time Hubble rate, the potential would be effectively negligible around the matter radiation equality and the mechanism of the model would be unaffected.
\item introduce two types of DM components (i.e. $\xi\not=1$) or a non vanishing $w_{\rm D}$. In that case, this offers the freedom to introduce a  small degree of isocurvature perturbations in the initial conditions in the dark sector;
\item one can consider that  $\tilde g_{\mu\nu}$ is not conformally related to $g_{\mu\nu}$ but through a disformal relation~\cite{Bekenstein:1992pj} as already considered to account for the dark sector~\cite{Zumalacarregui:2010wj}. 
\end{enumerate} 

So far we have defined and  exhibited models that improve slightly the Hubble tension. Indeed this is no proof that they describe nature and one would need to go deeper and exhibit some specific signatures. Since the theory is fully defined it opens the way to investigate its phenomenology in various environments. From a theoretical perspective, the model opens at least two concerns that would need to be investigated by specific DM theoretical constructions. (1) $\varphi$ is massless (or light) and unless protected by a symmetry, quantum corrections will generate a mass from DM field loops. This is the case for any interacting light fields. (2) One shall address the question of whether one can modify the DM gravity sector independently of the SM sector. In some models, a scalar fifth force may potentially lead to a violation of the WEP that could be probed~\cite{Carroll:2009dw,Mantry:2009ay}, in particular if DM and SM fields interact. Such models are however strongly constrained~\cite{Carroll:2008ub}. These two issues can only be discussed in a  model-dependent way and are far beyond the phenomenological investigation of the current work. Note also that it has also been suggested that the anomalies in the positron/electron spectra may arise from a long-range dark force mediating the DM annihilation and that it could be detected at the LHC~\cite{Bai:2009it}. Then, one shall investigate the effects of the scalar force in astrophysical environments and how it modifies DM haloes, see e.g. in Refs.~\cite{Frieman:1991zxc,Gradwohl:1992ue,Nusser:2004qu,Kesden:2006vz,Bean:2008ac,Keselman:2009nx,Mohapi:2015gua} for tests of the WEP between the visible and DM sectors. Our current analysis is limited to the linear regime. The age of the universe being smaller than in the standard $\Lambda$CDM and the growth of DM haloes being modified, one should quantify the onset of the non-linear regime and its consequence for structure formation. Still, as shown on Fig.~\ref{fig:Pkratio} the linear power spectrum is almost not affected compared to the  $\Lambda$CDM  and the scalar field being light does not cluster, which tend to make us think that modifications to non-linear effects would be small. The analysis in the case of extended quintessence (i.e. scalar-tensor quintessence) was first proposed in Refs.~\cite{Schimd:2004nq,Benabed:2001dm} to show the effects were below the percent level. Such an analysis is beyond the scope of this work and would deserve a specific study.  Ref.~\cite{Peebles:2012sm} also  suggested that a DM fifth force seems to have beneficial effects for DM distribution on small scales and that an extra evanescent component of matter with evolving mass and a fifth force large enough may be needed to reach a better understanding of early assembly of more nearly isolated protogalaxies.

As a conclusion, these new models are encouraging. They offer a  simple and minimal extension of the $\Lambda$CDM that decreases the Hubble tension while being compatible with all local experiments and BBN. It offers the possibility to be tested in other environments and the existence of such a dark fifth force may be a path toward a better understanding of DM~\cite{Moody:1984ba,Bovy:2008gh}.\\

\vskip1cm
\acknowledgements{We thank Nabila Aghanim, Karim Benabed, Jean-Fran\c{c}ois  Cardoso, Ali Rida Khalife and Agn\`es Fert\'e for discussions.}

\appendix
\section{Frame relations}\label{appB}

As explained in \S~\ref{secII1}, $g_{\mu\nu}$ is the SM-frame metric. For SM field, there is just one frame since the gravity theory is standard GR. For the DM sector, $g_{\mu\nu}$ corresponds to an Einstein frame metric with associated Jordan frame metric $\tilde g_{\mu\nu}$ defined in Eq.~(\ref{e.tildeg}). It is easily checked that, in the DM Jordan frame, Eq.~(\ref{FEq_3}) and~(\ref{e.motion}) can be rewritten as
\begin{equation}\label{JF1}
\tilde \nabla_\mu \tilde T_{\rm(DM)}^{\mu\nu} = 0
\end{equation}
\begin{eqnarray}\label{JF2}
\tilde\gamma^\mu = 0
\end{eqnarray}
(assuming $q=1$), where all the quantities are defined with respect to the metric~(\ref{e.tildeg}). This expresses the fact that DM fields are universally coupled to $\tilde g_{\mu\nu}$ so that the WEP is valid in this frame, leading to the usual conservation and geodesic equations. Since $ \tilde u^\nu  \tilde u_\nu = -1$ it follows that $ \tilde u^\nu=A^{-1} u^\mu$ and $\tilde T_{\mu\nu}=A^{-2} T_{\mu\nu}$ so that $\tilde\rho=A^{-4}\rho$.

Now assume that $g_{\mu\nu}$ defines the line element~(\ref{e.metpert}) $\dd s^2$ while $\tilde g_{\mu\nu}$ defines a similar line element $\dd \tilde s^2$ but with scale factor $\tilde a$ and potentials $\tilde\Phi, \tilde\Psi$. Since, $\dd \tilde s^2=A^{2}\dd s^2$ and $A(\varphi+\chi)=A(\varphi)[1+\alpha\chi]$, one concludes that
\begin{equation}\label{e.JF2}
\tilde a = aA, \quad
\tilde\Phi =\Phi +\alpha\chi,\quad
\tilde\Psi =\Psi -\alpha\chi.
\end{equation}
For the matter sector, the scaling gives
\begin{equation}\label{e.JF3}
\tilde \rho = A^{-4}\rho, \quad
\tilde\delta =\delta -4\alpha\chi,\quad
\tilde v = v.
\end{equation}
Now in the DM-Einstein frame, the equations of motion are derived from Eq.~(\ref{JF1}), that is from their the usual conservation equations. Hence $\dot{\tilde\rho}+3\tilde{\cal H}(\tilde \rho+\tilde P)=0$ directly gives Eq.~(\ref{e.c2}). At the perturbation level, $\tilde\delta' + \Delta\tilde v-3\tilde\Psi'=0$ and $\tilde v'+\tilde{\cal H}\tilde v =-\tilde\Phi$, which are the standard equations~(\ref{e.pert4}) and~(\ref{e.pert5}), which once expressed in the DM-Einstein frame reduce directly to equations~(\ref{pertDM1}) and~(\ref{pertDM2}). This a consistency check. This also explains the structure of the perturbation equations under the form~(\ref{pertDM1b}-\ref{pertDM2b}) since they are the standard perturbed conservation equations in the DM-Einstein frame. Indeed the Einstein equations need to be written for $\Phi$ and $\Psi$.

\section{Constraints in the $H_0-\Omega_{\rm m}$ plane}\label{app:DADH}

Let us consider a spatially Euclidean FL spacetime, filled with only matter and a cosmological constant, which is a good approximation in the matter era. The spacing of the baryon-photon acoustic oscillations provide a BAO standard ruler, $r_{\rm D}$, given by 
\begin{equation}\label{e.rD}
r_{\rm D} = \int_{z_{\rm D}}^\infty \frac{c_s(z)}{H(z)}\dd z
\end{equation}
with $z_{\rm D}$ the redshift of the drag epoch and $c_s$ the sound speed. It corresponds to the distance traveled by sound waves between the end of inflation and the baryon/photon decoupling after recombination. Typically $r_{\rm D}\sim 147$~Mpc. A measurement of radial BAO provides the ratio~\cite{eBOSS:2020yzd}
\begin{equation}
\theta_H(z) = \frac{r_{\rm D}}{D_H(z)} = r_{\rm D} H(z)\,.
\end{equation}
For a given $\Omega_{\rm m}$, the Hubble constant deduced from that measurement is
\begin{equation}
H_0 = \frac{\theta_H(z)}{r_{\rm D}}\frac{1}{\sqrt{\Omega_{\rm m}(1+z)^3 + 1- \Omega_{\rm m}}}\,.
\end{equation}
A measurement of angular BAO provides the ratio
\begin{equation}
\theta_A(z) = \frac{r_{\rm D}}{R_{\rm ang}(z)}\,.
\end{equation}
Since the universe is assumed to be spatially Euclidean, we have the relation~\eqref{eq:relDADH}, then for a given $\Omega_{\rm m}$ the Hubble constant deduced from that measurement is
\begin{equation}
H_0 = \frac{\theta_A(z)}{r_{\rm D}}\left[I(\Omega_{\rm m},1+z)-I(\Omega_{\rm m},1)\right]
\end{equation}
with
\begin{equation}
I(\Omega_{\rm m}, 1+z) = \frac{1+z}{\sqrt{1- \Omega_{\rm m}}}\, {}_2 F_1\left[\frac{1}{3},\frac{1}{2},\frac{4}{3},\frac{\Omega_{\rm m} (1+z)^3}{\Omega_{\rm m}-1} \right].\nonumber
\end{equation}

\bibliographystyle{apsrev4-1}
\bibliography{BIBHO}
\end{document}